\pdfoutput=1

\documentclass[preprint, prd, tightenlines, superscriptaddress]{revtex4-1}

\usepackage{graphicx}
\usepackage{dcolumn}
\usepackage{color}
\usepackage{multirow}
\usepackage{placeins}
\usepackage{amsmath}
\usepackage{amssymb}
\usepackage{amsfonts}
\usepackage{upgreek}
\usepackage{xspace}
\usepackage{url}
\usepackage{hyperref}
\usepackage[italic]{hepnames}
\usepackage{lineno}
\usepackage{siunitx}

\input{belle2sym.tex}
\newcommand{\CS} {\ensuremath{\text{CS}_{var}}\xspace}

\begin{document}

\vspace*{0.5\baselineskip}
\resizebox{!}{3cm}{\includegraphics{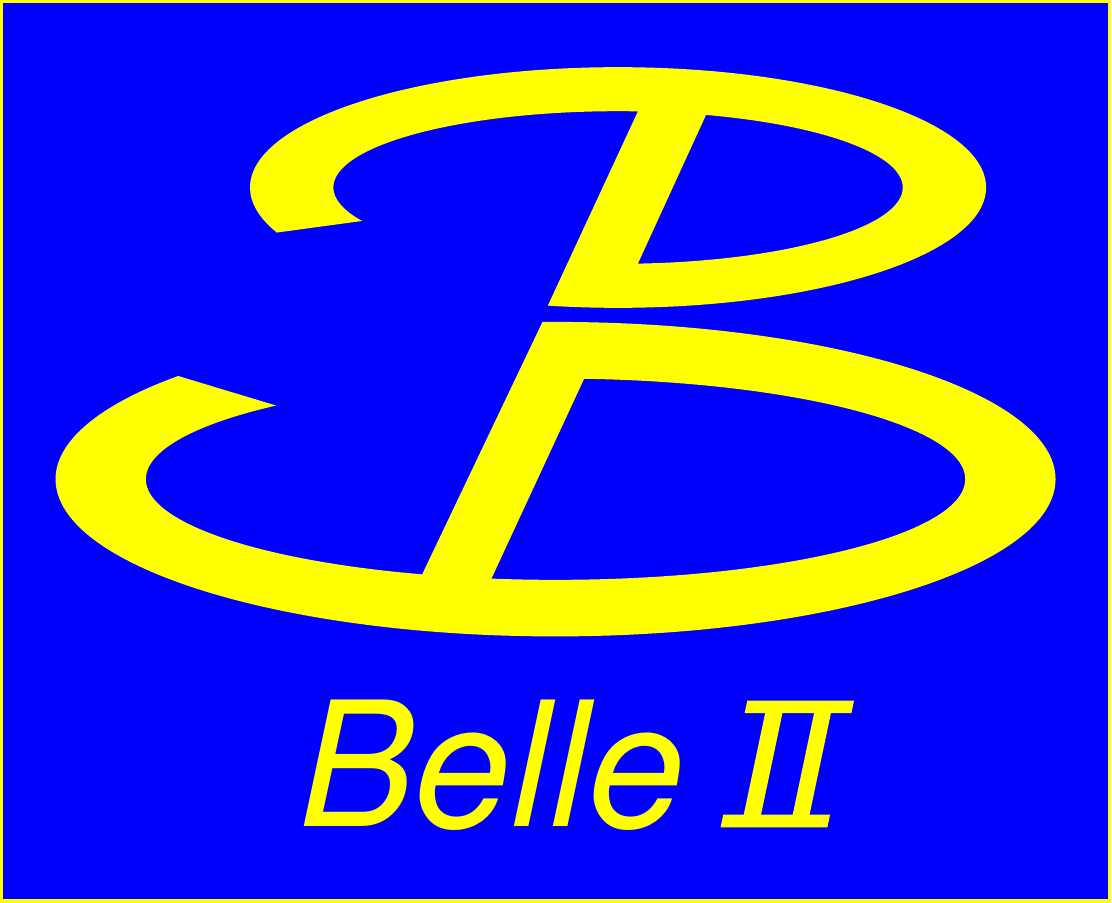}}

\vspace*{-4\baselineskip}
\begin{flushright}
BELLE2-CONF-PH-2021-005\\
\today  
\end{flushright}
\vspace*{2.5\baselineskip}

\title{Measurement of the branching fractions of $B\to\eta' K$ decays using 2019/2020 Belle II data}

\newcommand{\instCPPM}{Aix Marseille Universit\'{e}, CNRS/IN2P3, CPPM, 13288 Marseille, France}
\newcommand{\instBeihang}{Beihang University, Beijing 100191, China}
\newcommand{\instBNL}{Brookhaven National Laboratory, Upton, New York 11973, U.S.A.}
\newcommand{\instBINP}{Budker Institute of Nuclear Physics SB RAS, Novosibirsk 630090, Russian Federation}
\newcommand{\instCMU}{Carnegie Mellon University, Pittsburgh, Pennsylvania 15213, U.S.A.}
\newcommand{\instCinvestavIPN}{Centro de Investigacion y de Estudios Avanzados del Instituto Politecnico Nacional, Mexico City 07360, Mexico}
\newcommand{\instPrague}{Faculty of Mathematics and Physics, Charles University, 121 16 Prague, Czech Republic}
\newcommand{\instChiangMai}{Chiang Mai University, Chiang Mai 50202, Thailand}
\newcommand{\instChiba}{Chiba University, Chiba 263-8522, Japan}
\newcommand{\instChonnam}{Chonnam National University, Gwangju 61186, South Korea}
\newcommand{\instConacyt}{Consejo Nacional de Ciencia y Tecnolog\'{\i}a, Mexico City 03940, Mexico}
\newcommand{\instDESY}{Deutsches Elektronen--Synchrotron, 22607 Hamburg, Germany}
\newcommand{\instDuke}{Duke University, Durham, North Carolina 27708, U.S.A.}
\newcommand{\instITAR}{ Duy Tan University, Hanoi 100000, Vietnam}
\newcommand{\instRomaENEA}{ENEA Casaccia, I-00123 Roma, Italy}
\newcommand{\instEri}{Earthquake Research Institute, University of Tokyo, Tokyo 113-0032, Japan}
\newcommand{\instJuelich}{Forschungszentrum J\"{u}lich, 52425 J\"{u}lich, Germany}
\newcommand{\instFuJen}{Department of Physics, Fu Jen Catholic University, Taipei 24205, Taiwan}
\newcommand{\instFudan}{Key Laboratory of Nuclear Physics and Ion-beam Application (MOE) and Institute of Modern Physics, Fudan University, Shanghai 200443, China}
\newcommand{\instGoettingen}{II. Physikalisches Institut, Georg-August-Universit\"{a}t G\"{o}ttingen, 37073 G\"{o}ttingen, Germany}
\newcommand{\instGifu}{Gifu University, Gifu 501-1193, Japan}
\newcommand{\instSOKENDAI}{The Graduate University for Advanced Studies (SOKENDAI), Hayama 240-0193, Japan}
\newcommand{\instGyeongsang}{Gyeongsang National University, Jinju 52828, South Korea}
\newcommand{\instHanyang}{Department of Physics and Institute of Natural Sciences, Hanyang University, Seoul 04763, South Korea}
\newcommand{\instKEK}{High Energy Accelerator Research Organization (KEK), Tsukuba 305-0801, Japan}
\newcommand{\instJPARC}{J-PARC Branch, KEK Theory Center, High Energy Accelerator Research Organization (KEK), Tsukuba 305-0801, Japan}
\newcommand{\instHSE}{Higher School of Economics (HSE), Moscow 101000, Russian Federation}
\newcommand{\instIISER}{Indian Institute of Science Education and Research Mohali, SAS Nagar, 140306, India}
\newcommand{\instIITBhubaneswar}{Indian Institute of Technology Bhubaneswar, Satya Nagar 751007, India}
\newcommand{\instIITGuwahati}{Indian Institute of Technology Guwahati, Assam 781039, India}
\newcommand{\instIITHyderabad}{Indian Institute of Technology Hyderabad, Telangana 502285, India}
\newcommand{\instIITMadras}{Indian Institute of Technology Madras, Chennai 600036, India}
\newcommand{\instIndiana}{Indiana University, Bloomington, Indiana 47408, U.S.A.}
\newcommand{\instIHEPRussia}{Institute for High Energy Physics, Protvino 142281, Russian Federation}
\newcommand{\instHEPHYVienna}{Institute of High Energy Physics, Vienna 1050, Austria}
\newcommand{\instHiroshima}{Hiroshima University, Higashi-Hiroshima, Hiroshima 739-8530, Japan}
\newcommand{\instIHEPChina}{Institute of High Energy Physics, Chinese Academy of Sciences, Beijing 100049, China}
\newcommand{\instIPP}{Institute of Particle Physics (Canada), Victoria, British Columbia V8W 2Y2, Canada}
\newcommand{\instIOP}{Institute of Physics, Vietnam Academy of Science and Technology (VAST), Hanoi, Vietnam}
\newcommand{\instIFIC}{Instituto de Fisica Corpuscular, Paterna 46980, Spain}
\newcommand{\instFrascati}{INFN Laboratori Nazionali di Frascati, I-00044 Frascati, Italy}
\newcommand{\instNapoliINFN}{INFN Sezione di Napoli, I-80126 Napoli, Italy}
\newcommand{\instPadovaINFN}{INFN Sezione di Padova, I-35131 Padova, Italy}
\newcommand{\instPerugiaINFN}{INFN Sezione di Perugia, I-06123 Perugia, Italy}
\newcommand{\instPisaINFN}{INFN Sezione di Pisa, I-56127 Pisa, Italy}
\newcommand{\instRomaINFN}{INFN Sezione di Roma, I-00185 Roma, Italy}
\newcommand{\instRomaTreINFN}{INFN Sezione di Roma Tre, I-00146 Roma, Italy}
\newcommand{\instTorinoINFN}{INFN Sezione di Torino, I-10125 Torino, Italy}
\newcommand{\instTriesteINFN}{INFN Sezione di Trieste, I-34127 Trieste, Italy}
\newcommand{\instJAEA}{Advanced Science Research Center, Japan Atomic Energy Agency, Naka 319-1195, Japan}
\newcommand{\instMainz}{Johannes Gutenberg-Universit\"{a}t Mainz, Institut f\"{u}r Kernphysik, D-55099 Mainz, Germany}
\newcommand{\instGiessen}{Justus-Liebig-Universit\"{a}t Gie\ss{}en, 35392 Gie\ss{}en, Germany}
\newcommand{\instKarlsruhe}{Institut f\"{u}r Experimentelle Teilchenphysik, Karlsruher Institut f\"{u}r Technologie, 76131 Karlsruhe, Germany}
\newcommand{\instISU}{Iowa State University, Ames, Iowa 50011, U.S.A.}
\newcommand{\instKitasato}{Kitasato University, Sagamihara 252-0373, Japan}
\newcommand{\instKISTI}{Korea Institute of Science and Technology Information, Daejeon 34141, South Korea}
\newcommand{\instKoreaUnivKU}{Korea University, Seoul 02841, South Korea}
\newcommand{\instKSU}{Kyoto Sangyo University, Kyoto 603-8555, Japan}
\newcommand{\instKyungpook}{Kyungpook National University, Daegu 41566, South Korea}
\newcommand{\instLPI}{P.N. Lebedev Physical Institute of the Russian Academy of Sciences, Moscow 119991, Russian Federation}
\newcommand{\instLNNU}{Liaoning Normal University, Dalian 116029, China}
\newcommand{\instLMU}{Ludwig Maximilians University, 80539 Munich, Germany}
\newcommand{\instLuther}{Luther College, Decorah, Iowa 52101, U.S.A.}
\newcommand{\instMNITJaipur}{Malaviya National Institute of Technology Jaipur, Jaipur 302017, India}
\newcommand{\instMPP}{Max-Planck-Institut f\"{u}r Physik, 80805 M\"{u}nchen, Germany}
\newcommand{\instMPGHLL}{Semiconductor Laboratory of the Max Planck Society, 81739 M\"{u}nchen, Germany}
\newcommand{\instMcGill}{McGill University, Montr\'{e}al, Qu\'{e}bec, H3A 2T8, Canada}
\newcommand{\instMEPhI}{Moscow Physical Engineering Institute, Moscow 115409, Russian Federation}
\newcommand{\instNagoya}{Graduate School of Science, Nagoya University, Nagoya 464-8602, Japan}
\newcommand{\instNagoyaKMI}{Kobayashi-Maskawa Institute, Nagoya University, Nagoya 464-8602, Japan}
\newcommand{\instNagoyaIAR}{Institute for Advanced Research, Nagoya University, Nagoya 464-8602, Japan}
\newcommand{\instNaraWu}{Nara Women's University, Nara 630-8506, Japan}
\newcommand{\instNTUTaiwan}{Department of Physics, National Taiwan University, Taipei 10617, Taiwan}
\newcommand{\instNUUTaiwan}{National United University, Miao Li 36003, Taiwan}
\newcommand{\instKrakow}{H. Niewodniczanski Institute of Nuclear Physics, Krakow 31-342, Poland}
\newcommand{\instNiigata}{Niigata University, Niigata 950-2181, Japan}
\newcommand{\instNSU}{Novosibirsk State University, Novosibirsk 630090, Russian Federation}
\newcommand{\instOkinawa}{Okinawa Institute of Science and Technology, Okinawa 904-0495, Japan}
\newcommand{\instOsakaCity}{Osaka City University, Osaka 558-8585, Japan}
\newcommand{\instRCNP}{Research Center for Nuclear Physics, Osaka University, Osaka 567-0047, Japan}
\newcommand{\instPNNL}{Pacific Northwest National Laboratory, Richland, Washington 99352, U.S.A.}
\newcommand{\instPanjab}{Panjab University, Chandigarh 160014, India}
\newcommand{\instPanjabPAU}{Punjab Agricultural University, Ludhiana 141004, India}
\newcommand{\instRIKENMSL}{Meson Science Laboratory, Cluster for Pioneering Research, RIKEN, Saitama 351-0198, Japan}
\newcommand{\instSeoul}{Seoul National University, Seoul 08826, South Korea}
\newcommand{\instSPU}{Showa Pharmaceutical University, Tokyo 194-8543, Japan}
\newcommand{\instSoochow}{Soochow University, Suzhou 215006, China}
\newcommand{\instSoongsil}{Soongsil University, Seoul 06978, South Korea}
\newcommand{\instLjubljanaJSI}{J. Stefan Institute, 1000 Ljubljana, Slovenia}
\newcommand{\instKyiv}{Taras Shevchenko National Univ. of Kiev, Kiev, Ukraine}
\newcommand{\instTata}{Tata Institute of Fundamental Research, Mumbai 400005, India}
\newcommand{\instTUM}{Department of Physics, Technische Universit\"{a}t M\"{u}nchen, 85748 Garching, Germany}
\newcommand{\instTelAviv}{Tel Aviv University, School of Physics and Astronomy, Tel Aviv, 69978, Israel}
\newcommand{\instToho}{Toho University, Funabashi 274-8510, Japan}
\newcommand{\instTohoku}{Department of Physics, Tohoku University, Sendai 980-8578, Japan}
\newcommand{\instTitech}{Tokyo Institute of Technology, Tokyo 152-8550, Japan}
\newcommand{\instTokyoMetropolitan}{Tokyo Metropolitan University, Tokyo 192-0397, Japan}
\newcommand{\instUAS}{Universidad Autonoma de Sinaloa, Sinaloa 80000, Mexico}
\newcommand{\instNapoliUNIV}{Dipartimento di Scienze Fisiche, Universit\`{a} di Napoli Federico II, I-80126 Napoli, Italy}
\newcommand{\instPadovaUNIV}{Dipartimento di Fisica e Astronomia, Universit\`{a} di Padova, I-35131 Padova, Italy}
\newcommand{\instPerugiaUNIV}{Dipartimento di Fisica, Universit\`{a} di Perugia, I-06123 Perugia, Italy}
\newcommand{\instPisaUNIV}{Dipartimento di Fisica, Universit\`{a} di Pisa, I-56127 Pisa, Italy}
\newcommand{\instRomaTreUNIV}{Dipartimento di Matematica e Fisica, Universit\`{a} di Roma Tre, I-00146 Roma, Italy}
\newcommand{\instTorinoUNIV}{Dipartimento di Fisica, Universit\`{a} di Torino, I-10125 Torino, Italy}
\newcommand{\instTriesteUNIV}{Dipartimento di Fisica, Universit\`{a} di Trieste, I-34127 Trieste, Italy}
\newcommand{\instMontreal}{Universit\'{e} de Montr\'{e}al, Physique des Particules, Montr\'{e}al, Qu\'{e}bec, H3C 3J7, Canada}
\newcommand{\instIJCLab}{Universit\'{e} Paris-Saclay, CNRS/IN2P3, IJCLab, 91405 Orsay, France}
\newcommand{\instIPHC}{Universit\'{e} de Strasbourg, CNRS, IPHC, UMR 7178, 67037 Strasbourg, France}
\newcommand{\instAdelaide}{Department of Physics, University of Adelaide, Adelaide, South Australia 5005, Australia}
\newcommand{\instBonn}{University of Bonn, 53115 Bonn, Germany}
\newcommand{\instUBC}{University of British Columbia, Vancouver, British Columbia, V6T 1Z1, Canada}
\newcommand{\instCincinnati}{University of Cincinnati, Cincinnati, Ohio 45221, U.S.A.}
\newcommand{\instFlorida}{University of Florida, Gainesville, Florida 32611, U.S.A.}
\newcommand{\instHawaii}{University of Hawaii, Honolulu, Hawaii 96822, U.S.A.}
\newcommand{\instHeidelberg}{University of Heidelberg, 68131 Mannheim, Germany}
\newcommand{\instLjubljanaUniLJ}{Faculty of Mathematics and Physics, University of Ljubljana, 1000 Ljubljana, Slovenia}
\newcommand{\instLouisville}{University of Louisville, Louisville, Kentucky 40292, U.S.A.}
\newcommand{\instMalaya}{National Centre for Particle Physics, University Malaya, 50603 Kuala Lumpur, Malaysia}
\newcommand{\instLjubljanaUM}{University of Maribor, 2000 Maribor, Slovenia}
\newcommand{\instMelbourne}{School of Physics, University of Melbourne, Victoria 3010, Australia}
\newcommand{\instMississippi}{University of Mississippi, University, Mississippi 38677, U.S.A.}
\newcommand{\instUOM}{University of Miyazaki, Miyazaki 889-2192, Japan}
\newcommand{\instPittsburgh}{University of Pittsburgh, Pittsburgh, Pennsylvania 15260, U.S.A.}
\newcommand{\instUSTC}{University of Science and Technology of China, Hefei 230026, China}
\newcommand{\instSAlabama}{University of South Alabama, Mobile, Alabama 36688, U.S.A.}
\newcommand{\instSCarolina}{University of South Carolina, Columbia, South Carolina 29208, U.S.A.}
\newcommand{\instSydney}{School of Physics, University of Sydney, New South Wales 2006, Australia}
\newcommand{\instUTokyo}{Department of Physics, University of Tokyo, Tokyo 113-0033, Japan}
\newcommand{\instIPMU}{Kavli Institute for the Physics and Mathematics of the Universe (WPI), University of Tokyo, Kashiwa 277-8583, Japan}
\newcommand{\instVictoria}{University of Victoria, Victoria, British Columbia, V8W 3P6, Canada}
\newcommand{\instVPI}{Virginia Polytechnic Institute and State University, Blacksburg, Virginia 24061, U.S.A.}
\newcommand{\instWayneState}{Wayne State University, Detroit, Michigan 48202, U.S.A.}
\newcommand{\instYamagata}{Yamagata University, Yamagata 990-8560, Japan}
\newcommand{\instYerevan}{Alikhanyan National Science Laboratory, Yerevan 0036, Armenia}
\newcommand{\instYonsei}{Yonsei University, Seoul 03722, South Korea}
\newcommand{\instZZU}{Zhengzhou University, Zhengzhou 450001, China}
\affiliation{\instCPPM}
\affiliation{\instBeihang}
\affiliation{\instBNL}
\affiliation{\instBINP}
\affiliation{\instCMU}
\affiliation{\instCinvestavIPN}
\affiliation{\instPrague}
\affiliation{\instChiangMai}
\affiliation{\instChiba}
\affiliation{\instChonnam}
\affiliation{\instConacyt}
\affiliation{\instDESY}
\affiliation{\instDuke}
\affiliation{\instITAR}
\affiliation{\instRomaENEA}
\affiliation{\instEri}
\affiliation{\instJuelich}
\affiliation{\instFuJen}
\affiliation{\instFudan}
\affiliation{\instGoettingen}
\affiliation{\instGifu}
\affiliation{\instSOKENDAI}
\affiliation{\instGyeongsang}
\affiliation{\instHanyang}
\affiliation{\instKEK}
\affiliation{\instJPARC}
\affiliation{\instHSE}
\affiliation{\instIISER}
\affiliation{\instIITBhubaneswar}
\affiliation{\instIITGuwahati}
\affiliation{\instIITHyderabad}
\affiliation{\instIITMadras}
\affiliation{\instIndiana}
\affiliation{\instIHEPRussia}
\affiliation{\instHEPHYVienna}
\affiliation{\instHiroshima}
\affiliation{\instIHEPChina}
\affiliation{\instIPP}
\affiliation{\instIOP}
\affiliation{\instIFIC}
\affiliation{\instFrascati}
\affiliation{\instNapoliINFN}
\affiliation{\instPadovaINFN}
\affiliation{\instPerugiaINFN}
\affiliation{\instPisaINFN}
\affiliation{\instRomaINFN}
\affiliation{\instRomaTreINFN}
\affiliation{\instTorinoINFN}
\affiliation{\instTriesteINFN}
\affiliation{\instJAEA}
\affiliation{\instMainz}
\affiliation{\instGiessen}
\affiliation{\instKarlsruhe}
\affiliation{\instISU}
\affiliation{\instKitasato}
\affiliation{\instKISTI}
\affiliation{\instKoreaUnivKU}
\affiliation{\instKSU}
\affiliation{\instKyungpook}
\affiliation{\instLPI}
\affiliation{\instLNNU}
\affiliation{\instLMU}
\affiliation{\instLuther}
\affiliation{\instMNITJaipur}
\affiliation{\instMPP}
\affiliation{\instMPGHLL}
\affiliation{\instMcGill}
\affiliation{\instMEPhI}
\affiliation{\instNagoya}
\affiliation{\instNagoyaKMI}
\affiliation{\instNaraWu}
\affiliation{\instNTUTaiwan}
\affiliation{\instNUUTaiwan}
\affiliation{\instKrakow}
\affiliation{\instNiigata}
\affiliation{\instNSU}
\affiliation{\instOkinawa}
\affiliation{\instOsakaCity}
\affiliation{\instRCNP}
\affiliation{\instPNNL}
\affiliation{\instPanjab}
\affiliation{\instPanjabPAU}
\affiliation{\instRIKENMSL}
\affiliation{\instSeoul}
\affiliation{\instSPU}
\affiliation{\instSoochow}
\affiliation{\instSoongsil}
\affiliation{\instLjubljanaJSI}
\affiliation{\instKyiv}
\affiliation{\instTata}
\affiliation{\instTUM}
\affiliation{\instTelAviv}
\affiliation{\instToho}
\affiliation{\instTohoku}
\affiliation{\instTitech}
\affiliation{\instTokyoMetropolitan}
\affiliation{\instUAS}
\affiliation{\instNapoliUNIV}
\affiliation{\instPadovaUNIV}
\affiliation{\instPerugiaUNIV}
\affiliation{\instPisaUNIV}
\affiliation{\instRomaTreUNIV}
\affiliation{\instTorinoUNIV}
\affiliation{\instTriesteUNIV}
\affiliation{\instMontreal}
\affiliation{\instIJCLab}
\affiliation{\instIPHC}
\affiliation{\instAdelaide}
\affiliation{\instBonn}
\affiliation{\instUBC}
\affiliation{\instCincinnati}
\affiliation{\instFlorida}
\affiliation{\instHawaii}
\affiliation{\instHeidelberg}
\affiliation{\instLjubljanaUniLJ}
\affiliation{\instLouisville}
\affiliation{\instMalaya}
\affiliation{\instLjubljanaUM}
\affiliation{\instMelbourne}
\affiliation{\instMississippi}
\affiliation{\instUOM}
\affiliation{\instPittsburgh}
\affiliation{\instUSTC}
\affiliation{\instSAlabama}
\affiliation{\instSCarolina}
\affiliation{\instSydney}
\affiliation{\instUTokyo}
\affiliation{\instIPMU}
\affiliation{\instVictoria}
\affiliation{\instVPI}
\affiliation{\instWayneState}
\affiliation{\instYamagata}
\affiliation{\instYerevan}
\affiliation{\instYonsei}
\affiliation{\instZZU}
  \author{F.~Abudin{\'e}n}\affiliation{\instTriesteINFN} 
  \author{I.~Adachi}\affiliation{\instKEK}\affiliation{\instSOKENDAI} 
  \author{R.~Adak}\affiliation{\instFudan} 
  \author{K.~Adamczyk}\affiliation{\instKrakow} 
  \author{P.~Ahlburg}\affiliation{\instBonn} 
  \author{J.~K.~Ahn}\affiliation{\instKoreaUnivKU} 
  \author{H.~Aihara}\affiliation{\instUTokyo} 
  \author{N.~Akopov}\affiliation{\instYerevan} 
  \author{A.~Aloisio}\affiliation{\instNapoliUNIV}\affiliation{\instNapoliINFN} 
  \author{F.~Ameli}\affiliation{\instRomaINFN} 
% ...

\collaboration{Belle II Collaboration}

\vspace*{5mm}

\begin{abstract}


This note describes the rediscovery of $\PB\to\Petaprime\PK$ decays in Belle II
data, both in the charged and neutral final state: $\PBz\to\Petaprime\PKs$ and
$\PBpm\to\Petaprime\PKpm$. The $\Petaprime$ is searched for in two decay modes:
$\Petaprime\to\Peta\Pgpp\Pgpm$ with $\Peta\to\Pphoton\Pphoton$, and
$\Petaprime\to\Prho\Pphoton$.
The analysis uses data collected in 2019 and 2020 at the SuperKEKB asymmetric
\epem\ collider, with an integrated luminosity of $62.8\invfb$, corresponding to
$68.2$ million of \BB pairs produced.
The signal yield is obtained via an unbinned maximum likelihood fit to signal
sensitive variables, obtaining branching ratios:

$$\mathcal{B}\left(\PBpm\to\Petaprime\PKpm\right)  = \left(63.4~^{+3.4}_{-3.3}\,\stat\pm3.2\,\syst\right) \times10^{-6} $$
$$\mathcal{B}\left(\PBz\to\Petaprime\PKz\right)  = \left(59.9~^{+5.8}_{-5.5}\,\stat\pm2.9\,\syst\right) \times10^{-6} $$

\noindent{which} are consistent with world average.

\end{abstract}

\maketitle

\section{Introduction}
\label{sec:introduction}

Charmless hadronic \PB decays provide a rich ground for studying the mechanisms
of \PB meson decays and the phenomenon of CP violation. 
In particular, the decay $\PB \to \Petaprime \PK$  is a rare charmless hadronic
\PB decay, mediated via hadronic penguin diagram, which is particularly
sensitive to new physics in the hadronic loop.
The measurements of CP violation parameters using time dependent CP violation
techniques are the most precise for this kind of decay, thanks to the relatively
large branching fraction.
These measurements are also very clean from the theoretical point of view, thanks to the very
limited tree pollution~\cite{BENEKE2005143}.

The $\PB\to\Petaprime\PK$ decay was initially discovered by
CLEO~\cite{PhysRevLett.80.3710,PhysRevLett.85.520}.
The current best measurements of branching ratio \B 
were obtained by  Belle~\cite{Schumann:2006bg} and
BaBar~\cite{Aubert:2009yx}, using $386$ and $467$ million \BB pairs,
respectively.
The current Belle II integrated luminosity, collected at the \PUpsilonFourS
resonance, does not allow, yet, to improve these measurements, but the rediscovery
of these final states is an important benchmark to demonstrate the capability
of the Belle II detector.
These $\PB\to\Petaprime\PK$ decays are characterized by complicated final states, with charged and
neutral particles, and intermediate resonances. Moreover, they are affected by
a large contamination due to background coming both from {\em continuum}
$\Pelectron\Ppositron\to\Pquark\APquark$ ($\Pquark = \Pup,\Pdown, \Pstrange,
    \Pcharm$) events 
as well as from misreconstructed signal events ({\em self
        cross feed} (SxF)).  The continuum suppression is achieved by a
multivariate discriminator \CS, which is validated on off-resonance data.
The signal yield is extracted with a multidimensional maximum
likelihood fit, using as input variables :
$\Mbc=\sqrt{E^{*2}_{beam}c^4-p^{*2}_Bc^2}$, $\Delta{E}=E^{*}_B-E_{beam}$ 
    (where $(p,E)^{*}_B$ are momentum and energy of the candidate \PB computed
        in the center of mass system, and $E_{beam}=\sqrt{s}/2$),
and the output of the continuum suppression discriminator.

Both charged and neutral decays are measured. 
Two decay modes for \Petaprime are considered:
$\Petaprime\to\Peta(\to\gamma\gamma)\Pgpp\Pgpm$ and
$\Petaprime\to\Prho(\to\Pgpp\Pgpm)\Pphoton$, while only the $\PKs\to\Pgpp\Pgpm$
decay has been used.

\section{The Belle II detector and dataset}
\label{sec:dataset}

The Belle II detector is described in detail in Ref.~\cite{Abe:2010gxa}.
The detector has a cylindrical structure around the beam pipe, placed partially
inside a solenoidal superconducting magnet providing a 1.5T magnetic field.
The innermost sub-detector is the vertex detector (VXD), formed by two layers of
silicon pixel sensors and four layers of silicon strips, devoted to tracking
and vertexing. It is surrounded by a large central drift chamber (CDC), with
small cells and filled with a helium ethane mixture, which provide precise
measurement of momenta of charged tracks as well as particle identification via
energy loss measurement ($dE/dx$).
Two Cherenkov detectors provide additional particle identification: the Time of
Propagation (TOP) counter in the barrel region, and the Aerogel Ring Imaging
Cherenkov (ARICH) in the forward region.
The last detector inside the solenoid is the electromagnetic calorimeter (ECL),
based on CsI(Tl) crystals, dedicated to photon and electron identification and measurement.
The return yoke of the magnet is instrumented with scintillator strips and
resistive plate chambers, to provide measurements for \PKl mesons and muons (KLM).
The coordinate system is defined by the $z$ axis, corresponding to the solenoid
axis, and roughly oriented with the electron beam, the polar angle $\theta$
defined with respect the $z$ axis, and the azimuthal angle $\phi$.

The dataset used for this analysis was collected by Belle II in 2019 and
2020 at the SuperKEKB asymmetric energy \epem\ collider~\cite{Akai:2018mbz}.
The integrated luminosity collected at a centre-of-mass (CM) energy
corresponding to the \PUpsilonFourS resonance is $62.8\invfb$, with an
additional $9.2\invfb$ collected about 60\mev below the resonance
(off-resonance dataset). This corresponds to $n(\PB\PaB)=68.21\times{}10^6$ produced
pairs of \BB~\cite{Collaboration:1997}.

\section{Event selection and continuum suppression}
\label{sec:selection}


Charged tracks are required to have their point of closest approach within 2 cm
(0.5 cm) of the measured \epem\ interaction point along the $z$ axis (in the
    transverse plane), and to be inside the CDC acceptance region.
Photons candidate clusters are required to have a minimum energy of 150\mev, more than 1.5
ECL cells in the cluster, and 
$E_9/E_{21}>0.9$, where $E_9$ is the energy of the $3\times3$ ECL
cells around the photon hit point and $E_{21}$ is the energy in the $5\times5$
cells excluding the outermost 4 cells at the vertices.

For the $\Petaprime\to\Peta(\to\gamma\gamma)\Pgpp\Pgpm$ decay mode, the
intermediate \Peta resonance is formed with two photons with an invariant mass
$0.5<M_{\Pphoton\Pphoton}<0.57\gevcc$; the \Petaprime is built from an
\Peta candidate and two oppositely charged tracks having a pion mass hypothesis, and with 
$0.92<M_{\Peta\Pgpp\Pgpm}<1.0\gevcc$.

For the $\Petaprime\to\Prho\Pphoton$ decay mode, the \Prho candidate also uses two
oppositely charged tracks each having a pion mass hypothesis, with 
at least 20 measurements in the CDC, and with a minimal requirement on particle
identification (PID$>0.1$), combining information
from all sub-detectors, for a typical efficiency of 90\%. The invariant mass of
the pion pair is required to be $0.51<M_{\Pgpp\Pgpm}<1.0\gevcc$.
The additional \Pphoton used to form the \Petaprime candidate is required to
have $\cos\theta>-0.64$, corresponding to the barrel and forward region of ECL. The
\Petaprime invariant mass is required to be $0.92<M_{\Prho\Pphoton}<1.0\gevcc$

For the charged final state $\PBpm\to\Petaprime\PKpm$, the $\PKpm$ is required to
have a minimal PID$>0.1$, more than 20 measurements in the CDC, and not lie in the backward part
of the acceptance: $\cos\theta_{\PKpm}>-0.5$.

The \PKs candidate used for the neutral final state is formed from two charged
tracks, with a pion mass hypothesis, with an invariant mass
$0.49<M_{\Pgpp\Pgpm}<0.51\gevcc$ and the cosine of the angle $\alpha_{p,v}$, between the direction of the
\Pgpp\Pgpm system and the direction defined by the \epem\ interaction point and
the \PKs reconstructed secondary vertex, is required to be $\cos\alpha_{p,v}>0.99$.

The full decay chain is reconstructed imposing constraints on the mass of the
intermediate resonances, with the exception of \Prho, to be the world average~\cite{Zyla:2020zbs}.

The candidate multiplicity is around two candidates per event for the
$\Petaprime\to\Peta\Pgpp\Pgpm$ decay channel, and about six for 
$\Petaprime\to\Prho\Pphoton$.
Only the candidate with the best \PB vertex probability is retained: the simulation
shows that this selects the correct candidate in about 95\% of the cases.
The overall efficiency for this selection is about $31\%$ and $24\%$ for the
$\Petaprime\to\Peta\Pgpp\Pgpm$ and $\Petaprime\to\Prho\Pphoton$ decay channels,
respectively. The self cross feed is about 2-3\% of the signal.

\begin{figure}[!htb]
    \includegraphics[width=.45\textwidth]{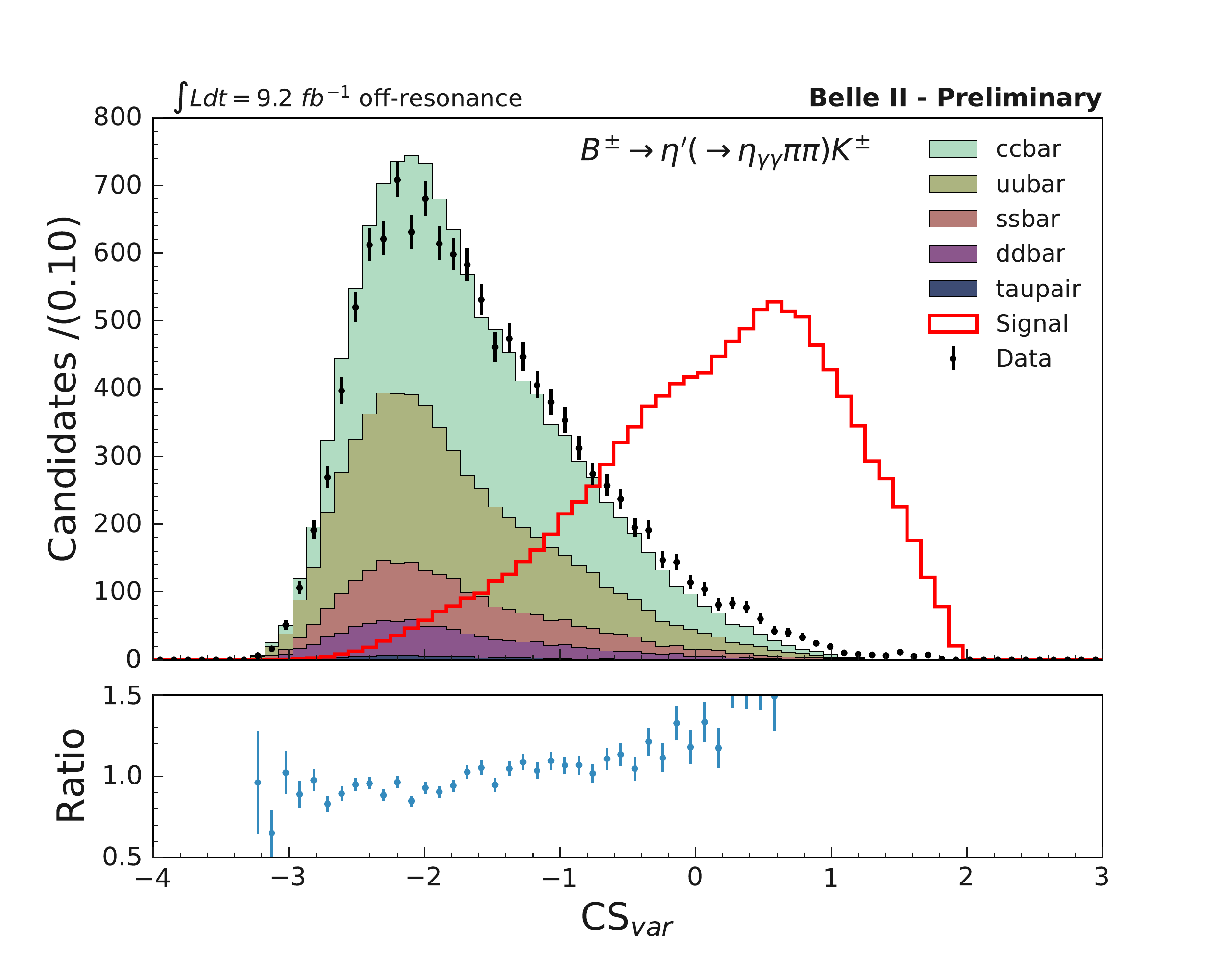}
    \includegraphics[width=.45\textwidth]{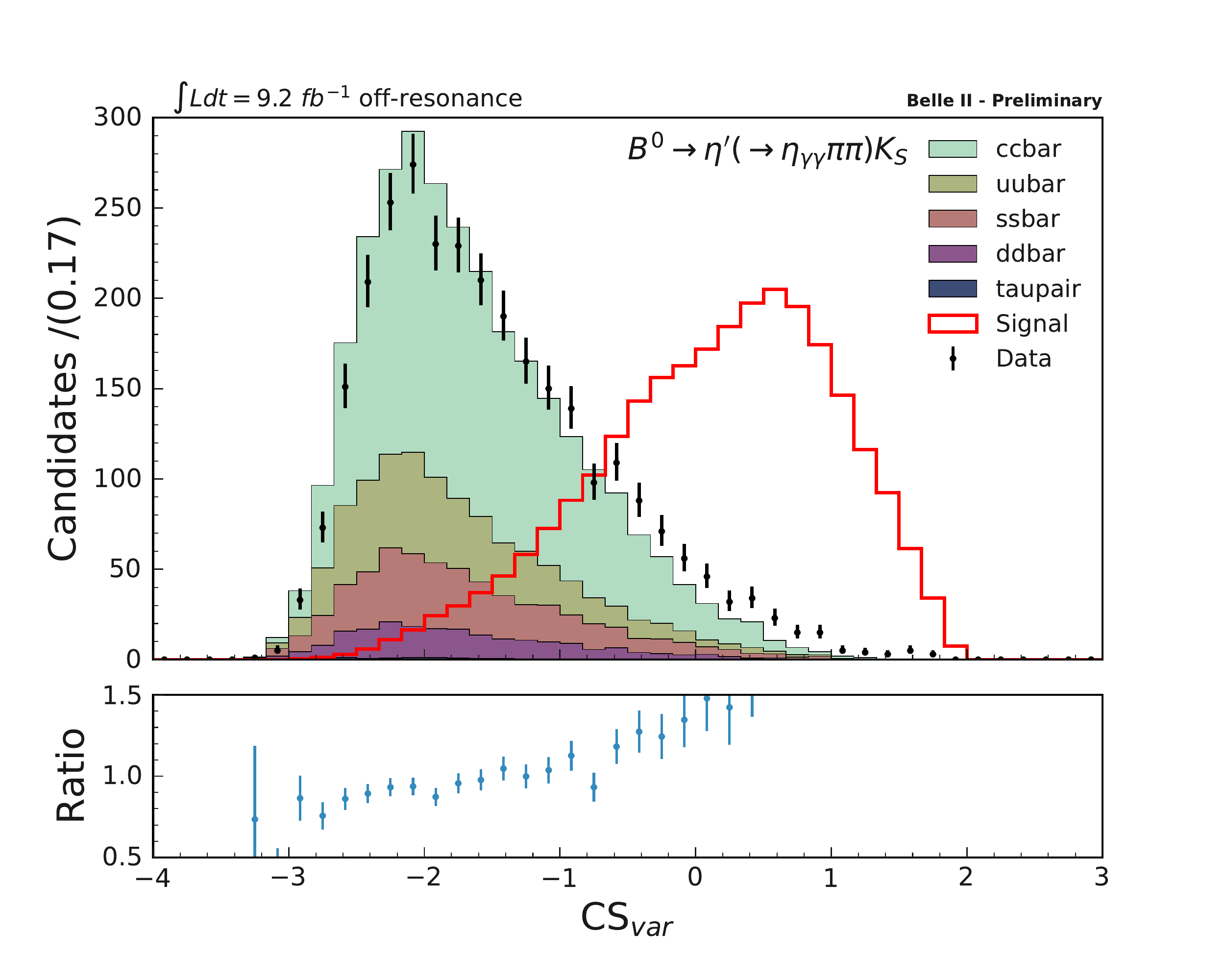}

    \includegraphics[width=.45\textwidth]{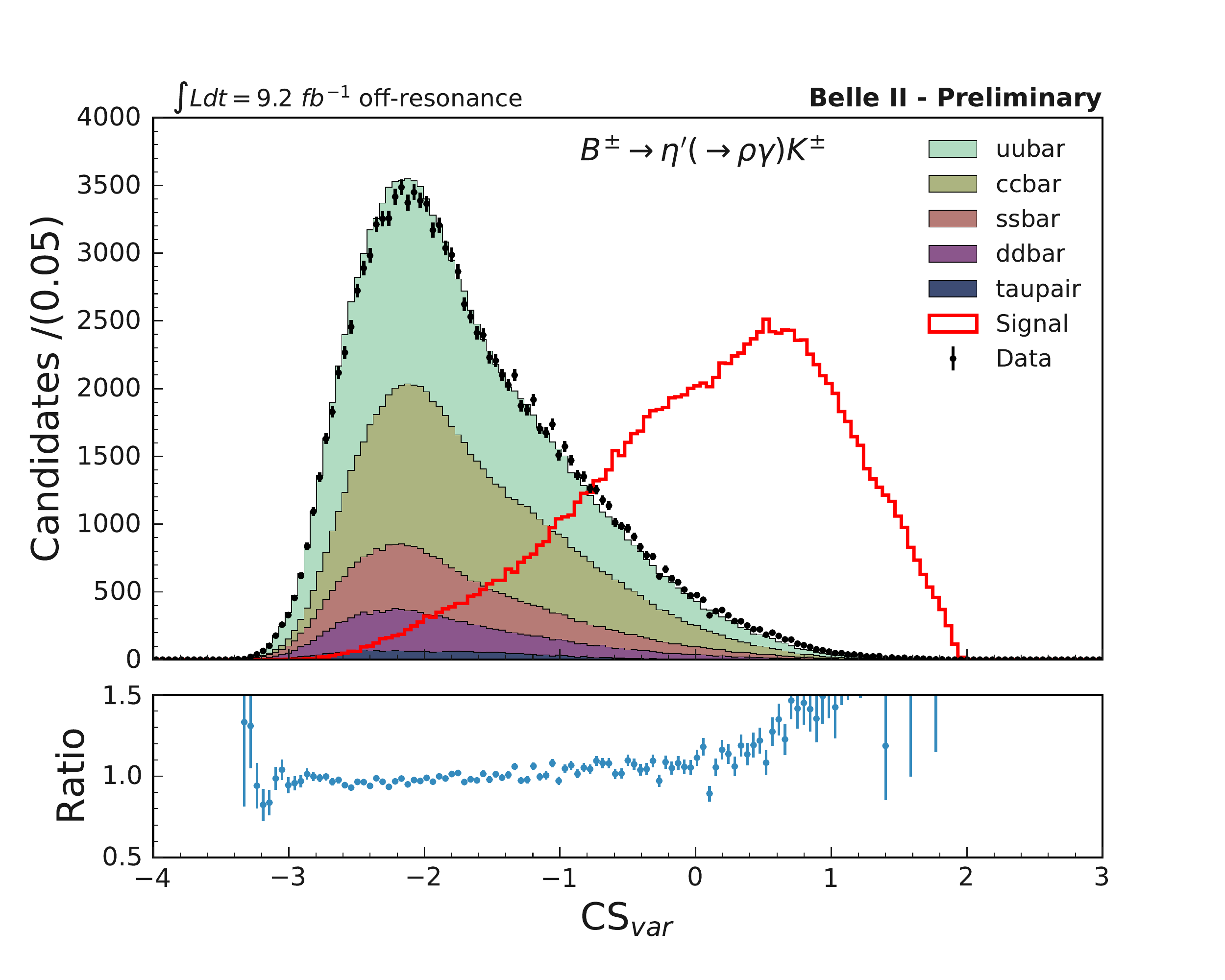}
    \includegraphics[width=.45\textwidth]{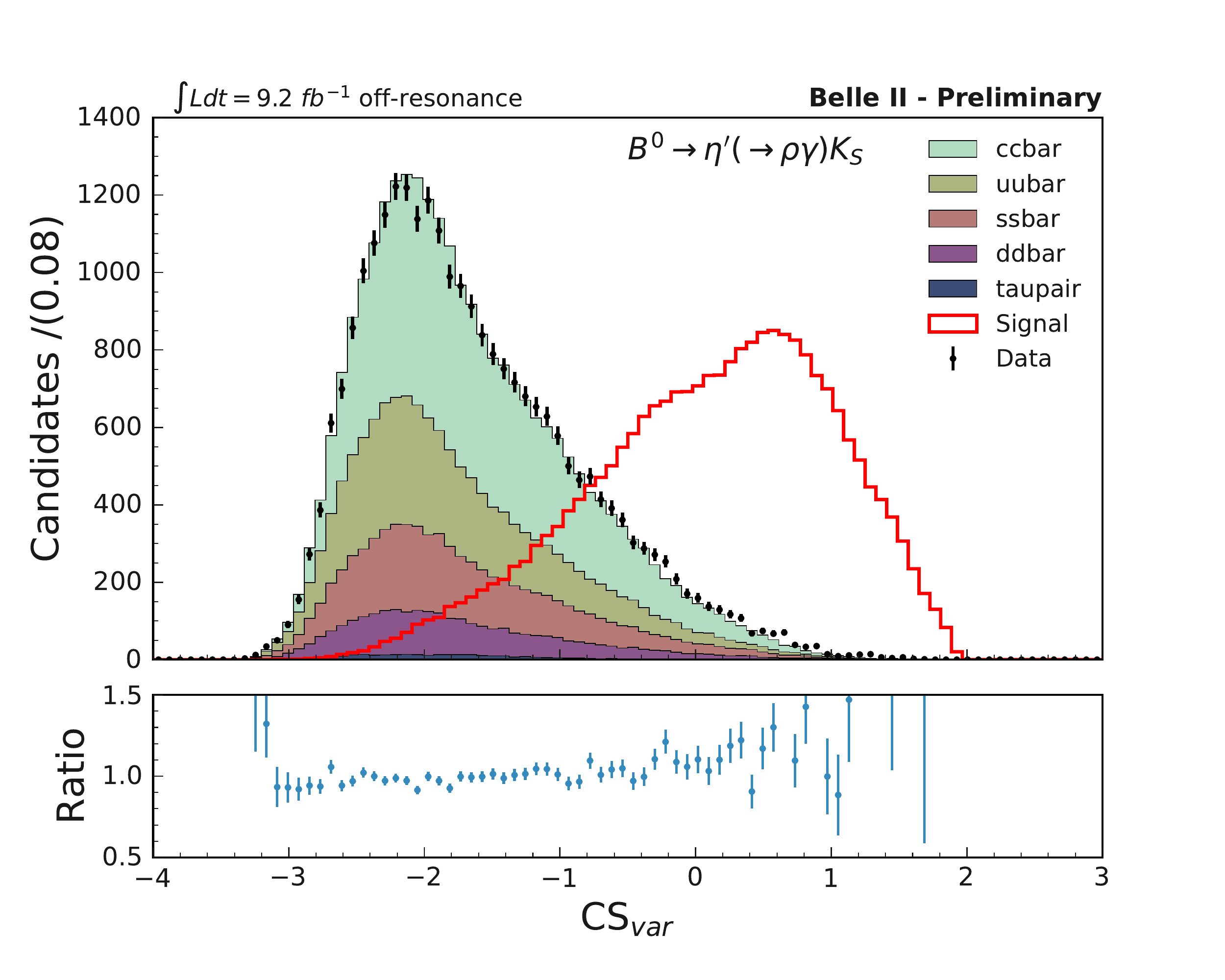}

    \caption{Distribution of continuum suppression multivariate discriminator for
        off-resonance data and continuum MC, normalized to same area, for the
        four decay channels, after signal selection.
        The signal distribution (in red), also normalized to same area, is superimposed.}
       
    \label{fig:orCSsig}
\end{figure}

To reduce the dominant background from random combinations of particles in
continuum events, we use a multivariate approach, combining a set of variables
which are sensitive to the event shape. These includes Kakuno-Super-Fox-Wolfram 
moments~\cite{Bevan:2014iga}, {\sc CLEO} cones~\cite{Asner:1995hc}, as well
as the angles of the thrust axis of signal \PB with respect to that the rest of event and that with respect the beam
axis. All variables that exhibit a correlation greater than 10\% with \Mbc and
\De are excluded. The classifier (\CS) used is based on the FastBDT algorithm~\cite{Keck:2017gsv}.
The output of the discriminator is validated using
off-resonance data, as shown in Fig.~\ref{fig:orCSsig}, where the
discrimination against signal is also visible.
The discrepancy in shape between off-resonance data and simulation is covered
by a dedicated systematic uncertainty.
No selection is applied to this variable, but it is directly used in the
fit to extract the signal yield, as described below.

\section{Maximum likelihood fit}
\label{sec:MLfit}

The signal extraction is performed via an extended, unbinned multivariate
maximum likelihood (ML) fit.  The likelihood $\mathcal{L}$ is defined, for the
$i^{th}$ events, with a set of observables $\vec{x}_i$, as:

\begin{equation}
    \mathcal{L}_i(\vec{x})=\sum_{j=1}^{m}n_j\mathcal{P}_j(\vec{x}_i)\quad,
\end{equation}

\noindent{where} $\mathcal{P}_j$ is the probability density for the component $j$ computed for
input variables (observables) $\vec{x}_i$, and $n_j$ is the number of events in
the dataset for component $j$.
The input observables $\vec{x}_i$ are $\Mbc$, $\Delta{E}$, and the continuum
suppression discriminator \CS.

The probabilities $\mathcal{P}_j$ are assumed to be the product of 1-dimensional
probability density functions ({\tt pdf}s) for each input variable, neglecting
correlations.
The correlation of the observables has been tested on simulation, where a small
(anti-)correlation ($\sim10\%$) between \Mbc and \De is observed, while the continuum
suppression discriminator shows no correlation with the other two variables.

For a dataset of $N$ events, where $N$ is expected to fluctuate according to
Poisson statistics, the likelihood $\mathcal{L}$ is:

\begin{equation}
    \mathcal{L}(N; \vec{x})=\frac{e^{-\sum n_j}}{N!}\prod_{i=1}^{N}\mathcal{L}_i \quad,
\end{equation}

The components considered for the fit are:
\begin{itemize}
    \item {\bf signal}: correctly reconstructed signal;
    \item {\bf signal cross feed (SxF)}: incorrectly reconstructed signal, namely
    events where a signal is present but where the reconstruction fails to assign
    the correct particles to form a \PB;
    \item {\bf continuum}: background from
    $\Pelectron\Ppositron\to\Pquark\APquark$ ($\Pquark = \Pup,\Pdown,
        \Pstrange, \Pcharm$) and $\Pelectron\Ppositron\to\Ptau\Ptau$;
    \item {\bf peaking}: background from \BB, both charged and neutral.
\end{itemize}

As the signal cross feed component is expected to be small and the observables
are not sensitive to it, the fit is performed considering signal and signal
cross feed as a single component, and the relative fraction (about 9\%) is
fixed from simulation.

The functional form of the {\tt pdf}s for all components are modelled from simulation.
We use a double Gaussian for \Mbc and \De for signal, for SxF a Crystal Ball~\cite{Oreglia:1980cs}
(CB) for \Mbc and CB or Gaussian plus polynomial for \De.
For continuum, an {\sc ARGUS} function~\cite{ALBRECHT1990278} is used for \Mbc, and a polynomial for \De.
For peaking background, the sum of an {\sc ARGUS} and a Gaussian is used for \Mbc and a polynomial is used for \De.
Finally, a bifurcated Gaussian is used for \CS.

The fit to data is performed by varying the yield of signal,
continuum, and peaking background, as well as the mean and width of the
Gaussian used to model \Mbc for signal, the slope of the {\sc ARGUS}
for continuum for \Mbc, the mean and width of the Gaussian for \De for signal,
and the parameters for the polynomial model for continuum \De.

The minimization is performed using {\sc MINUT}~\cite{James:1994vla}, and the
uncertainty computation uses the {\sc MINOS} algorithm.
The fit was validated and confirmed to be unbiased with
pseudo-experiments.
The events in the signal region (defined as $\Mbc>5.27\gevcc$ and
    $-0.07<\Delta{E}<0.05\gev$) were not examined until the fit procedure
was defined and validated, using simulation, side bands, and off-resonance
data.

\section{Results}
\label{sec:results}

\begin{figure}[!htb]
    \includegraphics[width=.48\textwidth]{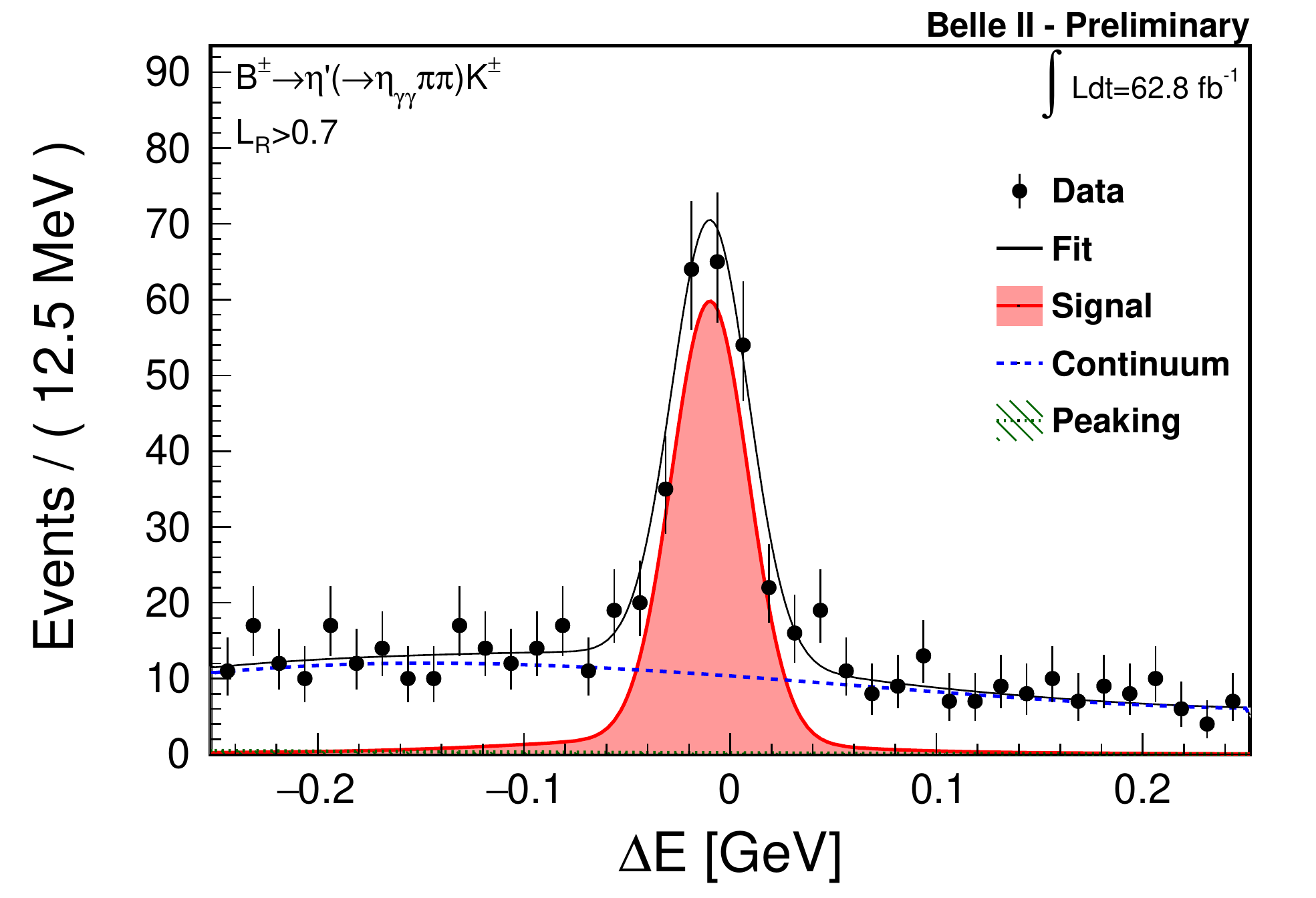}
    \includegraphics[width=.48\textwidth]{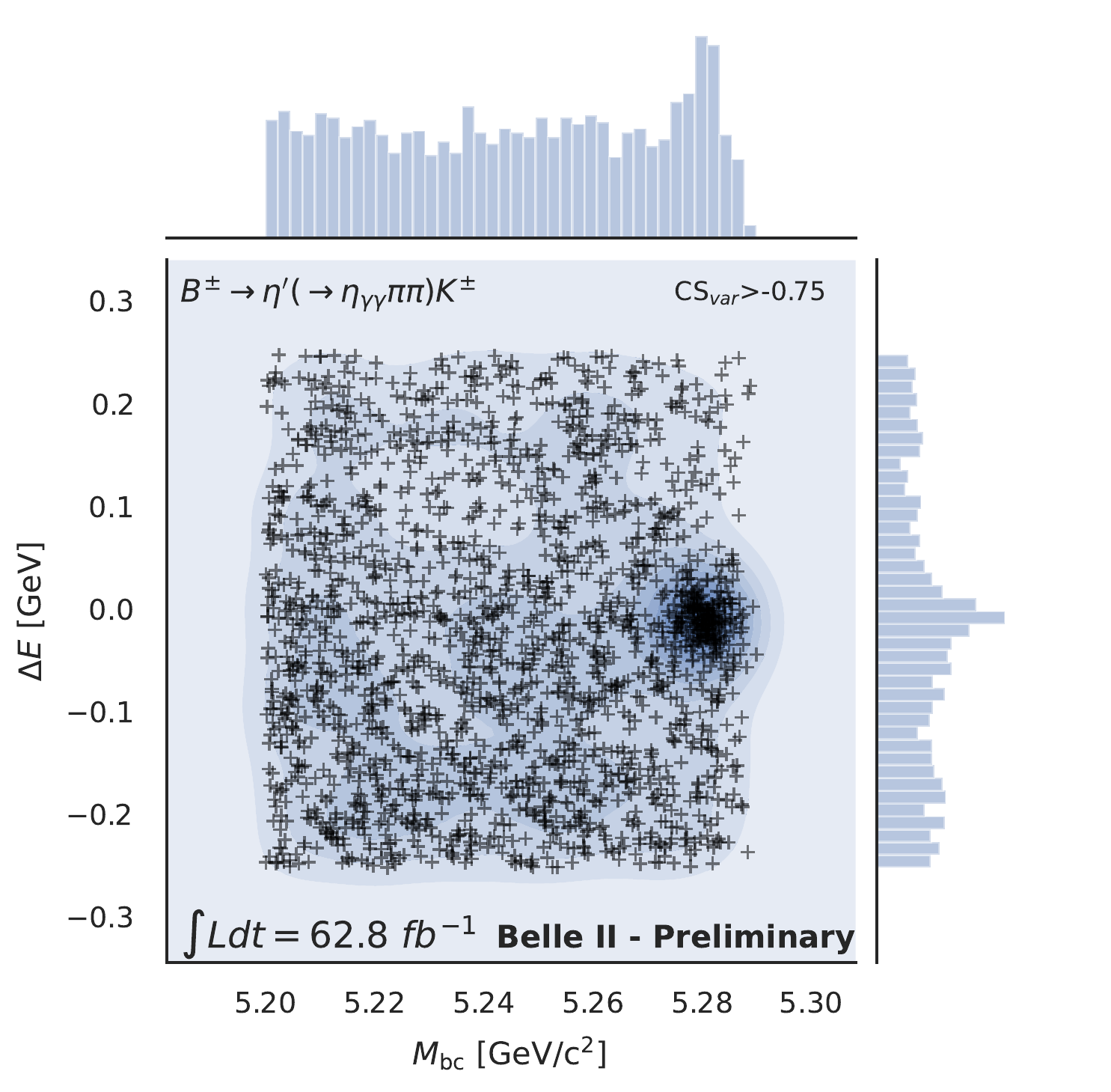}

    \includegraphics[width=.48\textwidth]{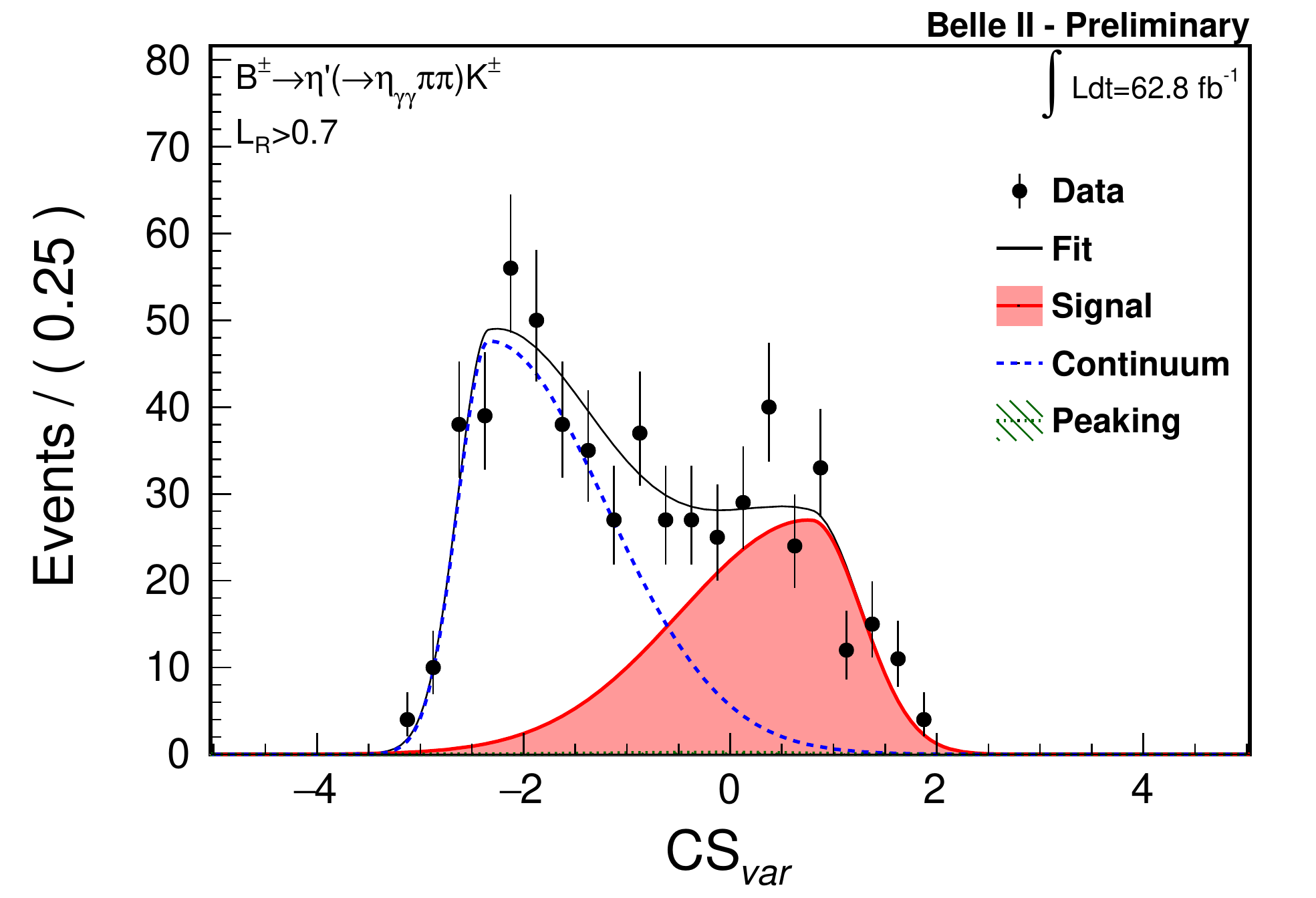}
    \includegraphics[width=.48\textwidth]{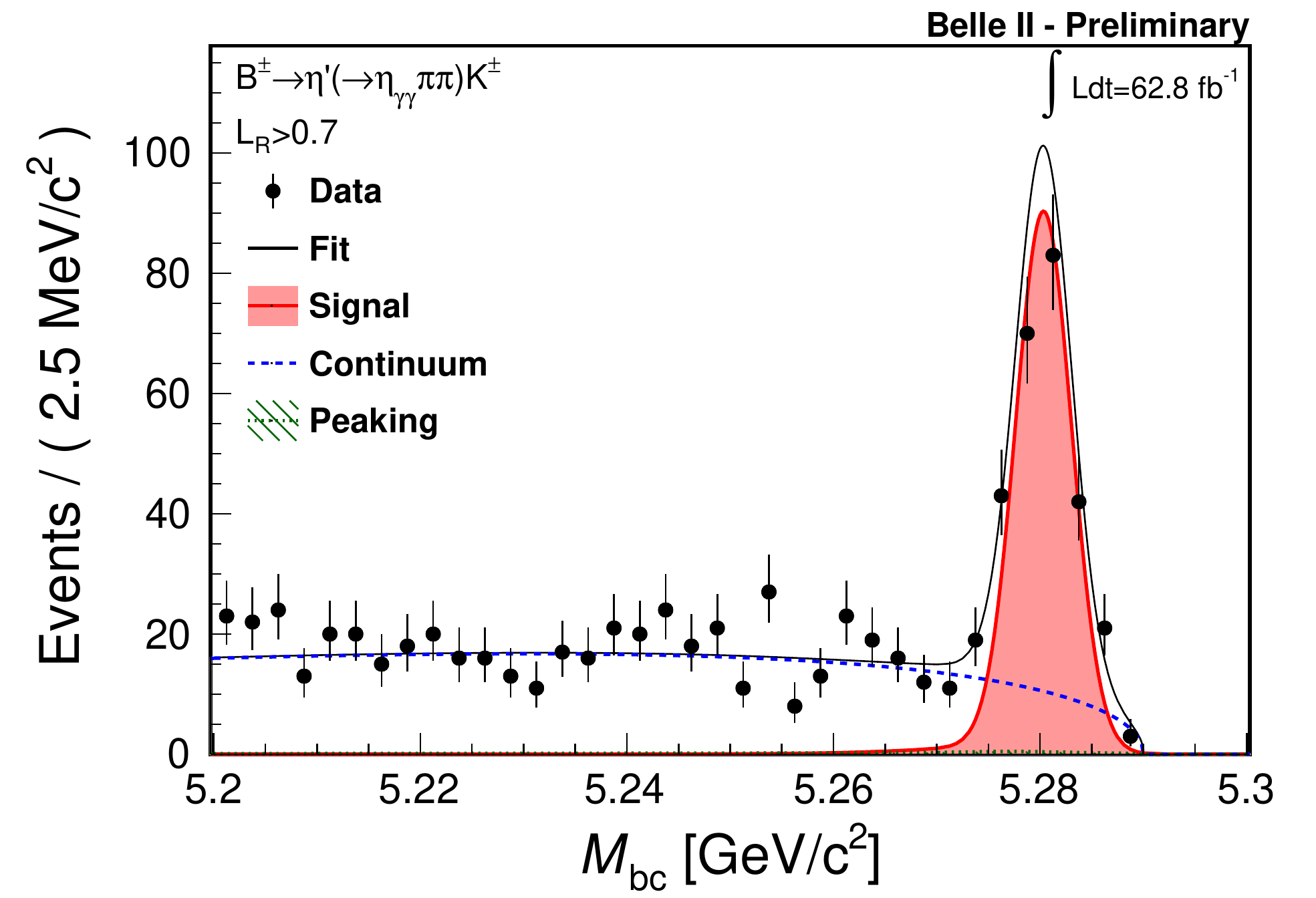}

    \caption{Distributions of $\Mbc$ and $\Delta{E}$, and continuum suppression discriminator for
        the signal-enriched region ($\mathcal{L}_R>0.7$), as well as $\Mbc$ versus $\Delta{E}$
        with the FoM-optimized \CS selection reported on the plot, for the channel 
        $\PBpm\to\Petaprime\PKpm$ with $\Petaprime\to\Peta\Pgpp\Pgpm$.
        Superimposed on the 1D distributions are
        the results of the extended ML fit as described in the text.
    }
       
    \label{fig:res_Bpch1}
\end{figure}

\begin{figure}[!htb]
    \includegraphics[width=.48\textwidth]{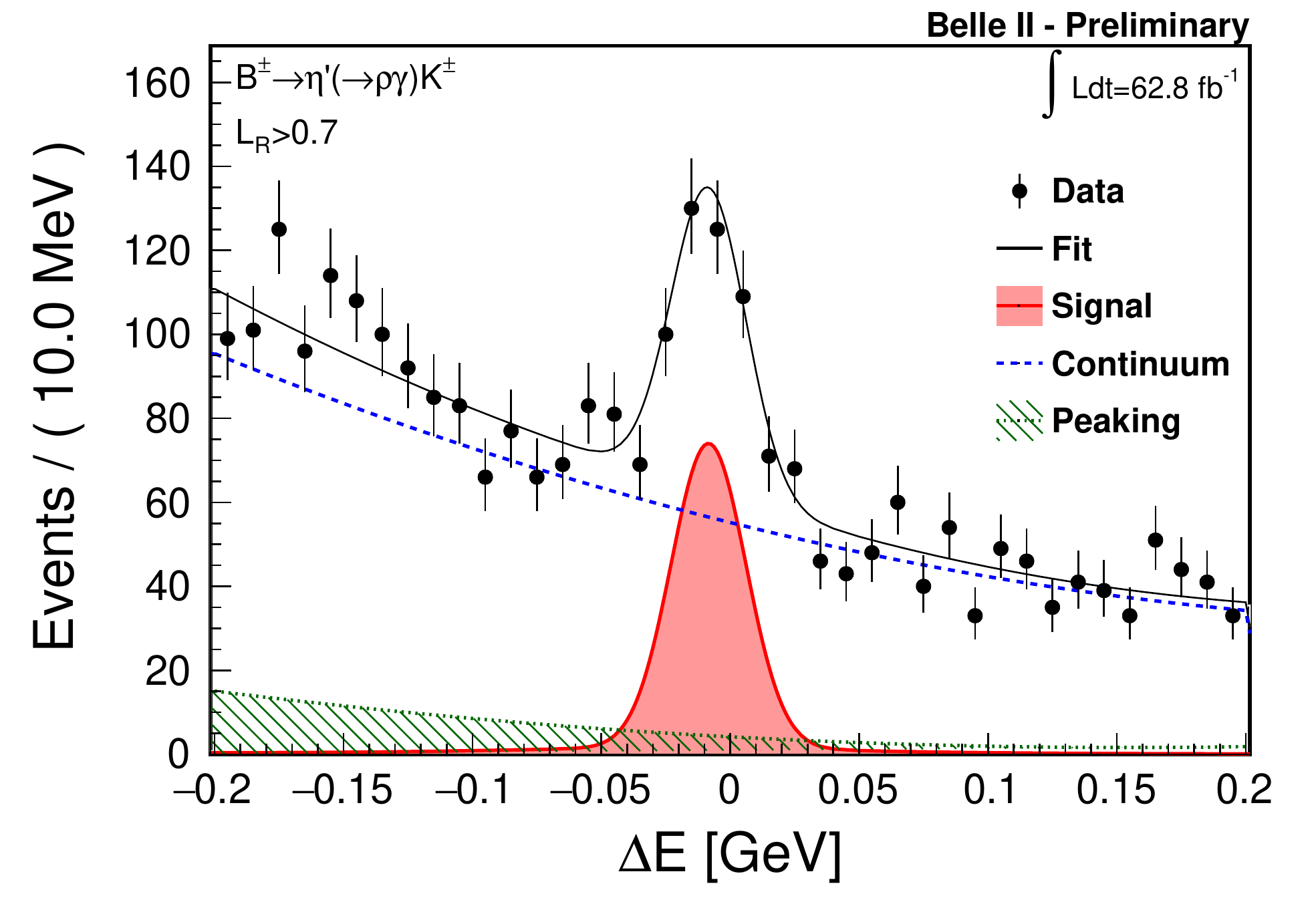}
    \includegraphics[width=.48\textwidth]{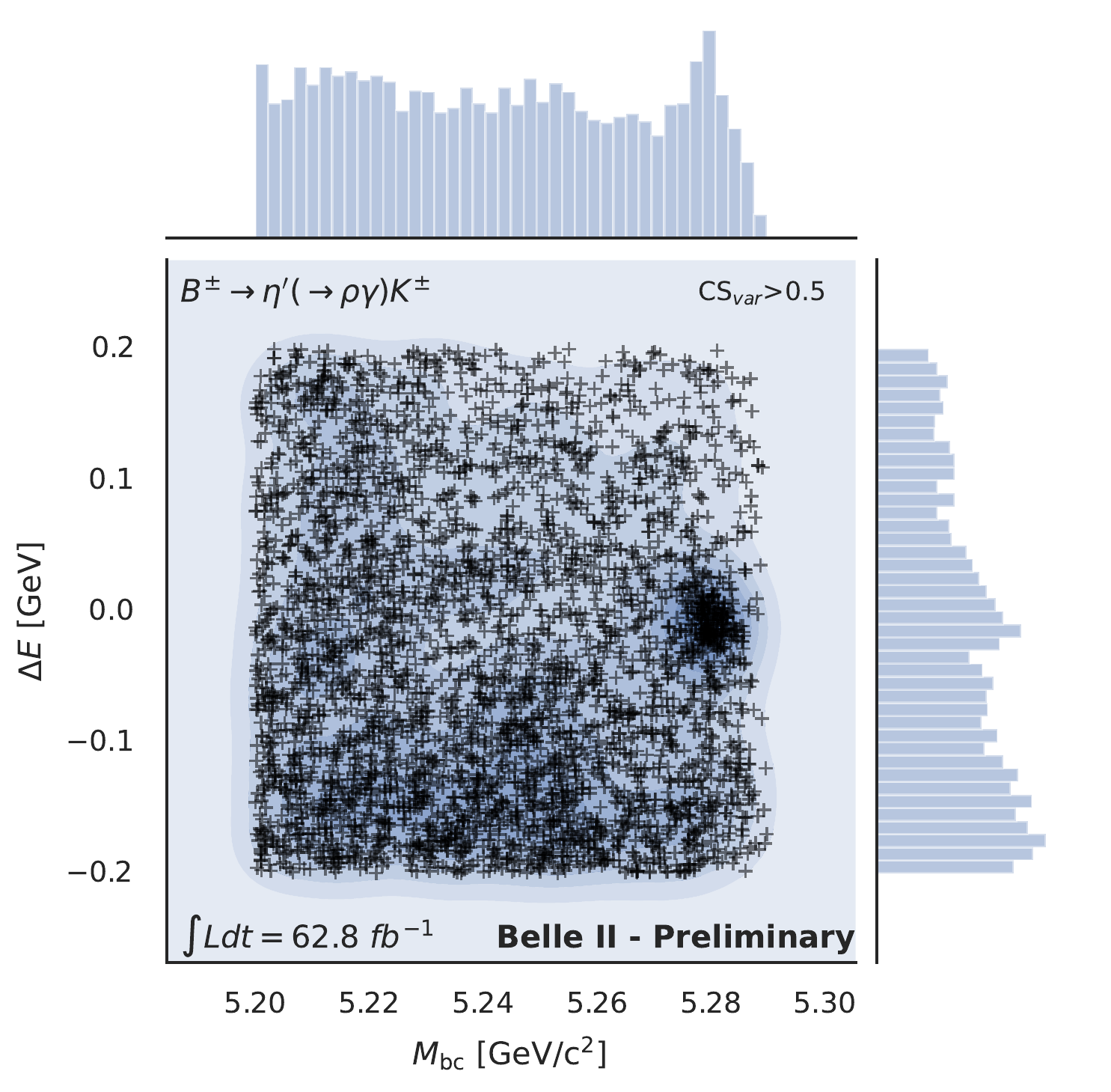}

    \includegraphics[width=.48\textwidth]{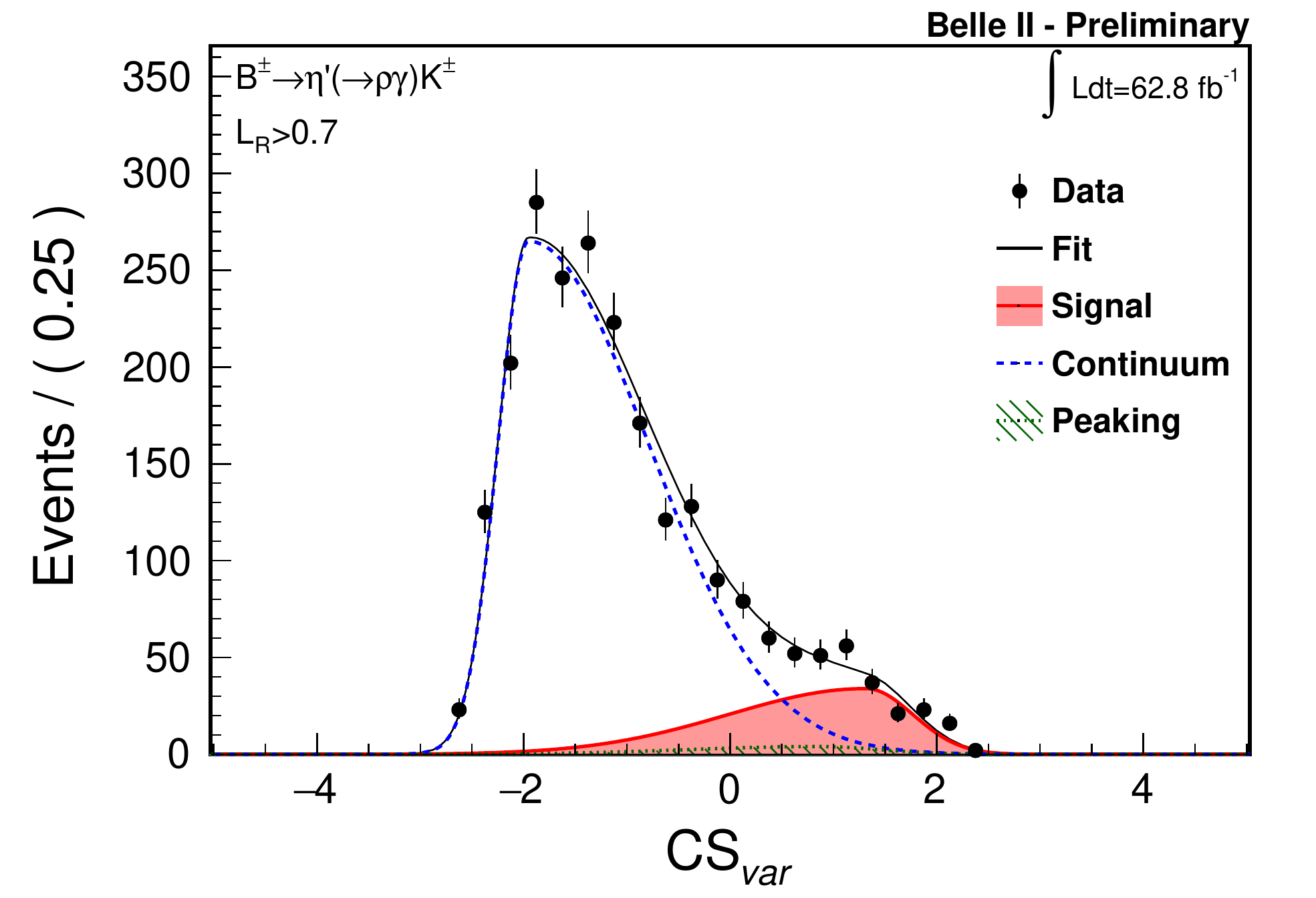}
    \includegraphics[width=.48\textwidth]{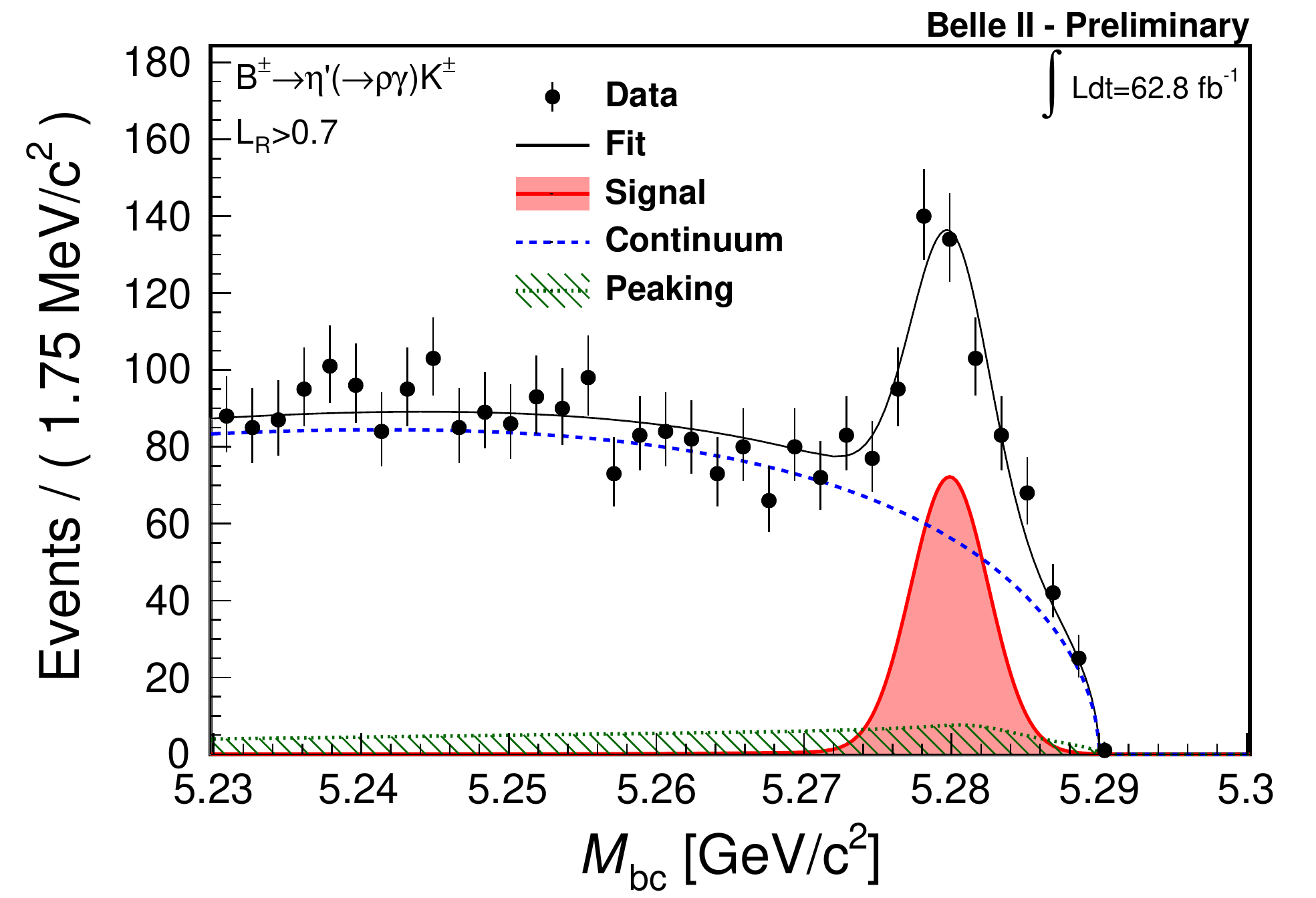}

    \caption{Distributions of $\Mbc$ and $\Delta{E}$, and continuum suppression discriminator for
        the signal-enriched region ($\mathcal{L}_R>0.7$), as well as $\Mbc$ versus $\Delta{E}$
        with the FoM-optimized \CS selection reported on the plot, for the channel 
        $\Petaprime\to\Prho\Pphoton$. Superimposed on the 1D distributions are
        the results of the extended ML fit as described in the text.
    }
       
    \label{fig:res_Bpch3}
\end{figure}

\begin{figure}[!htb]
    \includegraphics[width=.48\textwidth]{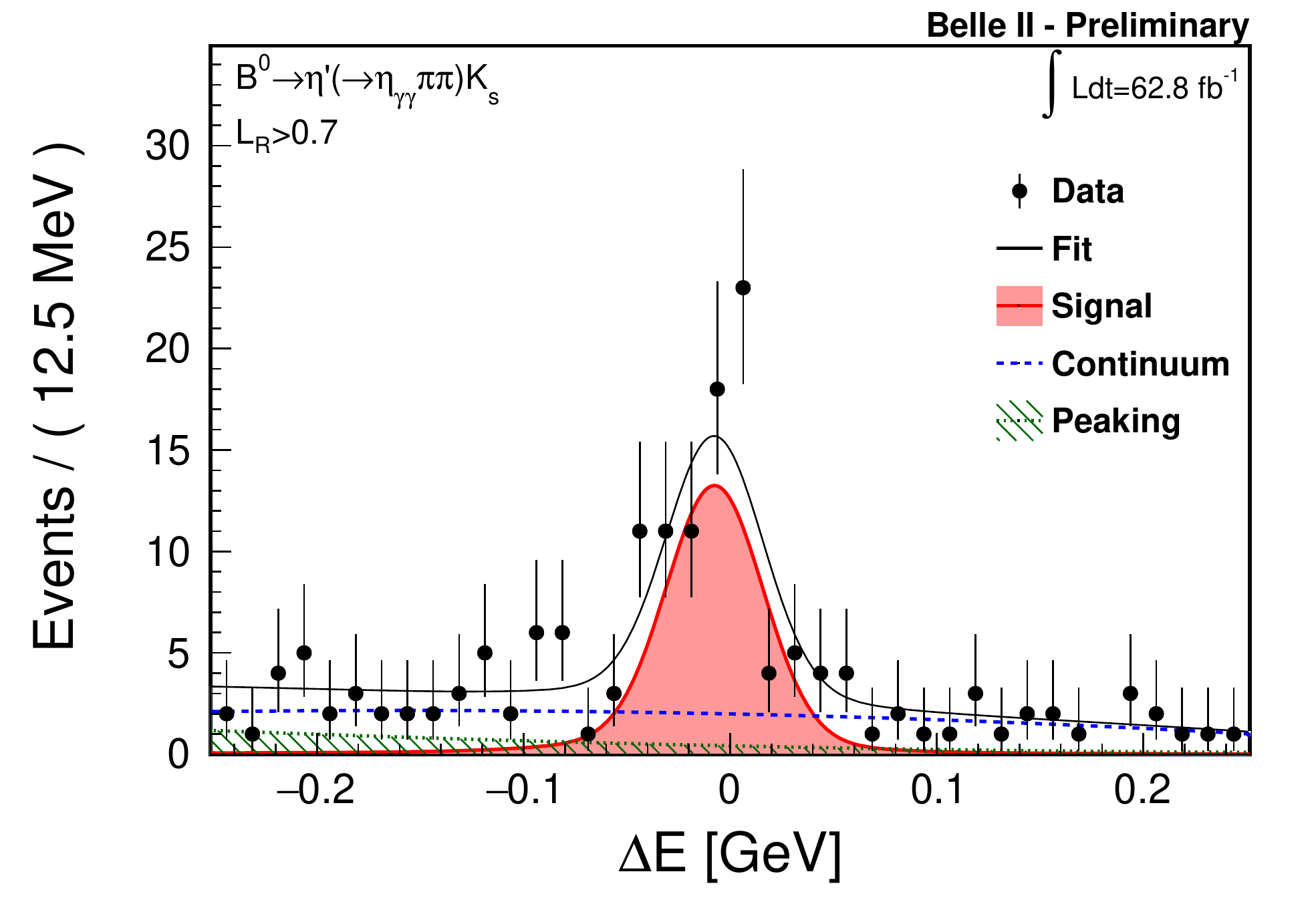}
    \includegraphics[width=.48\textwidth]{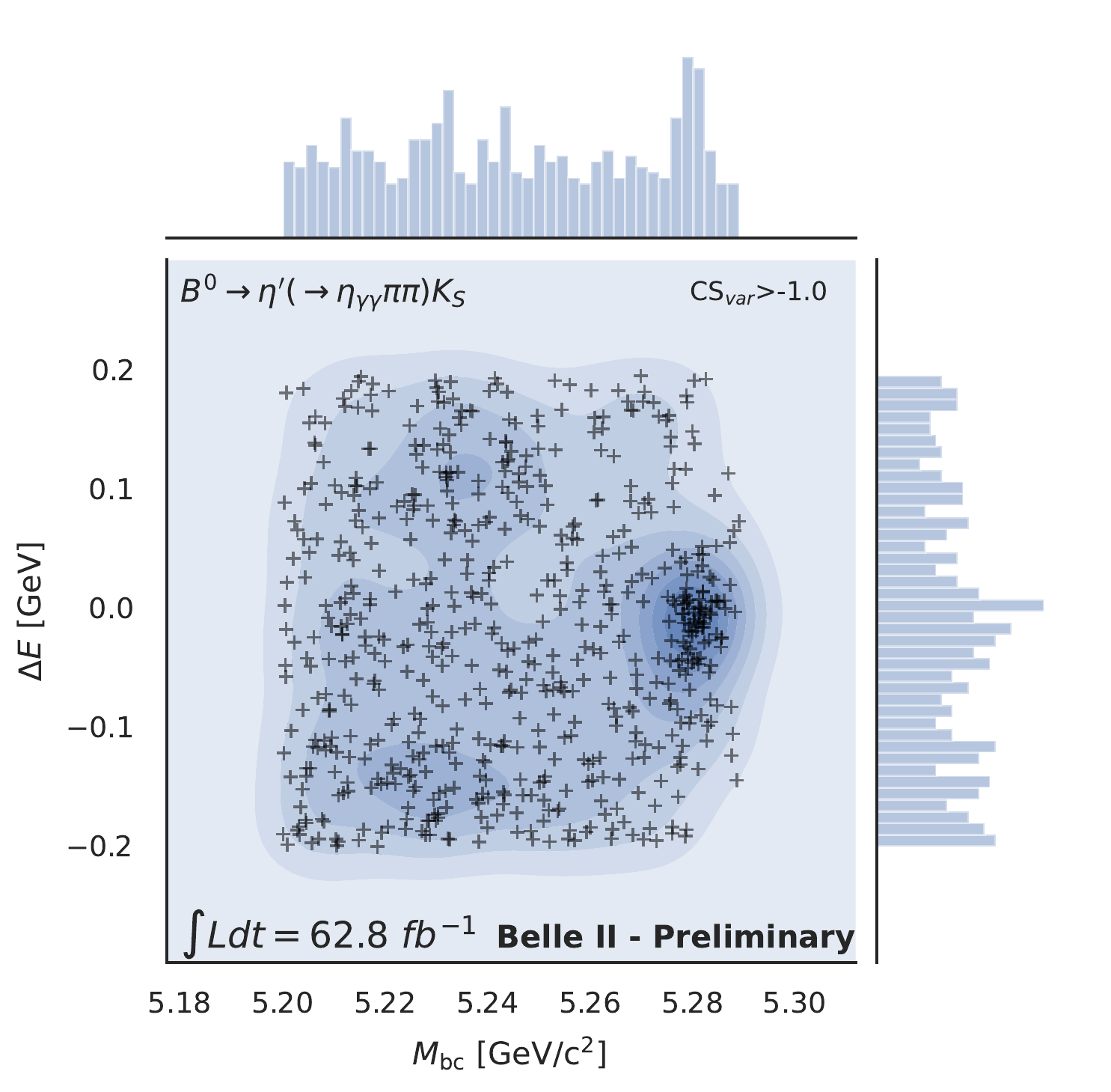}

    \includegraphics[width=.48\textwidth]{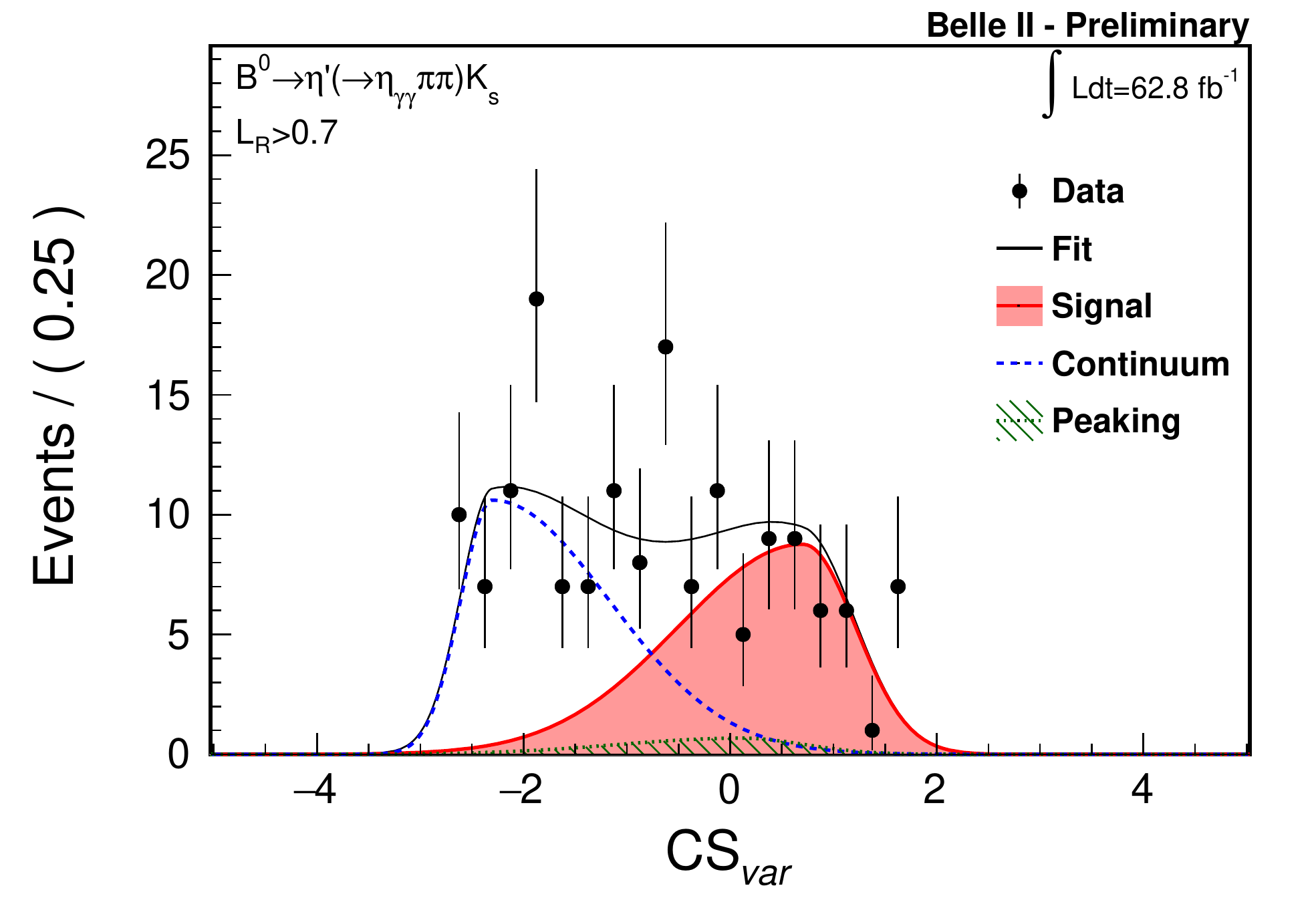}
    \includegraphics[width=.48\textwidth]{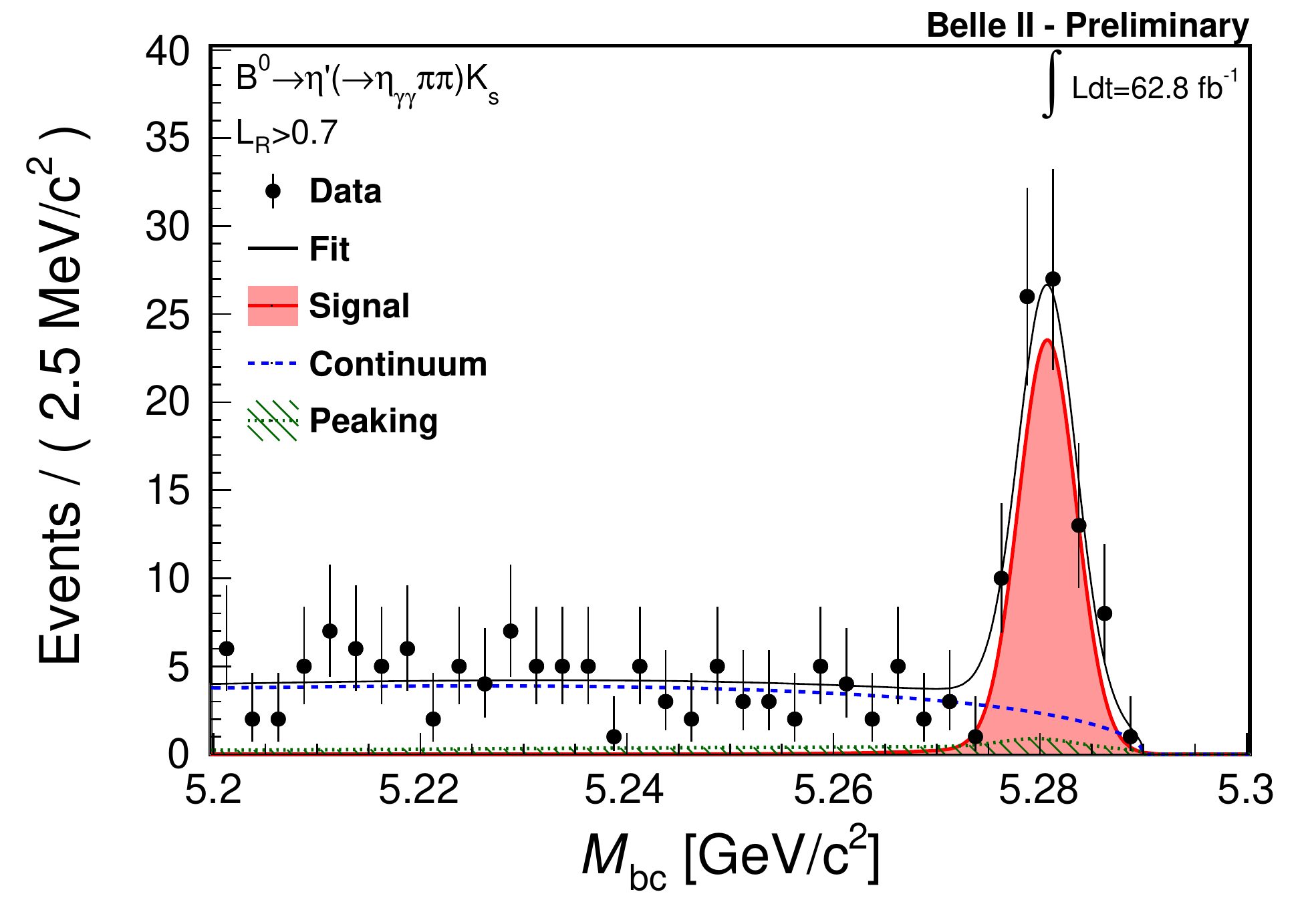}

    \caption{Distributions of $\Mbc$ and $\Delta{E}$, and continuum suppression discriminator for
        the signal-enriched region ($\mathcal{L}_R>0.7$), as well as $\Mbc$ versus $\Delta{E}$
        with the FoM-optimized \CS selection reported on the plot, for the channel 
        $\PBz\to\Petaprime\PKs$ with $\Petaprime\to\Peta\Pgpp\Pgpm$.
        Superimposed on the 1D distributions are
        the results of the extended ML fit as described in the text.
    }
       
    \label{fig:res_B0ch1}
\end{figure}

\begin{figure}[!htb]
    \includegraphics[width=.48\textwidth]{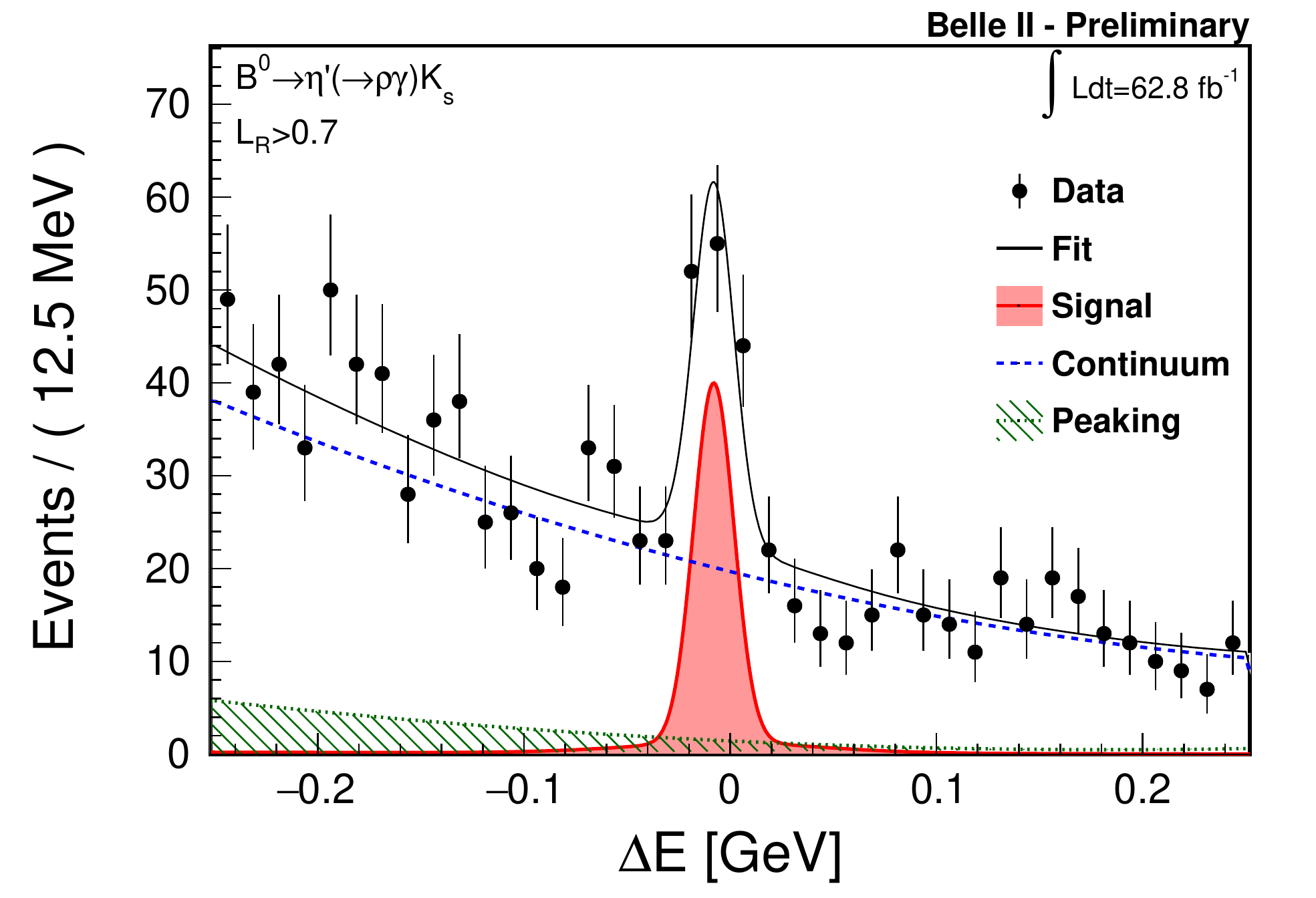}
    \includegraphics[width=.48\textwidth]{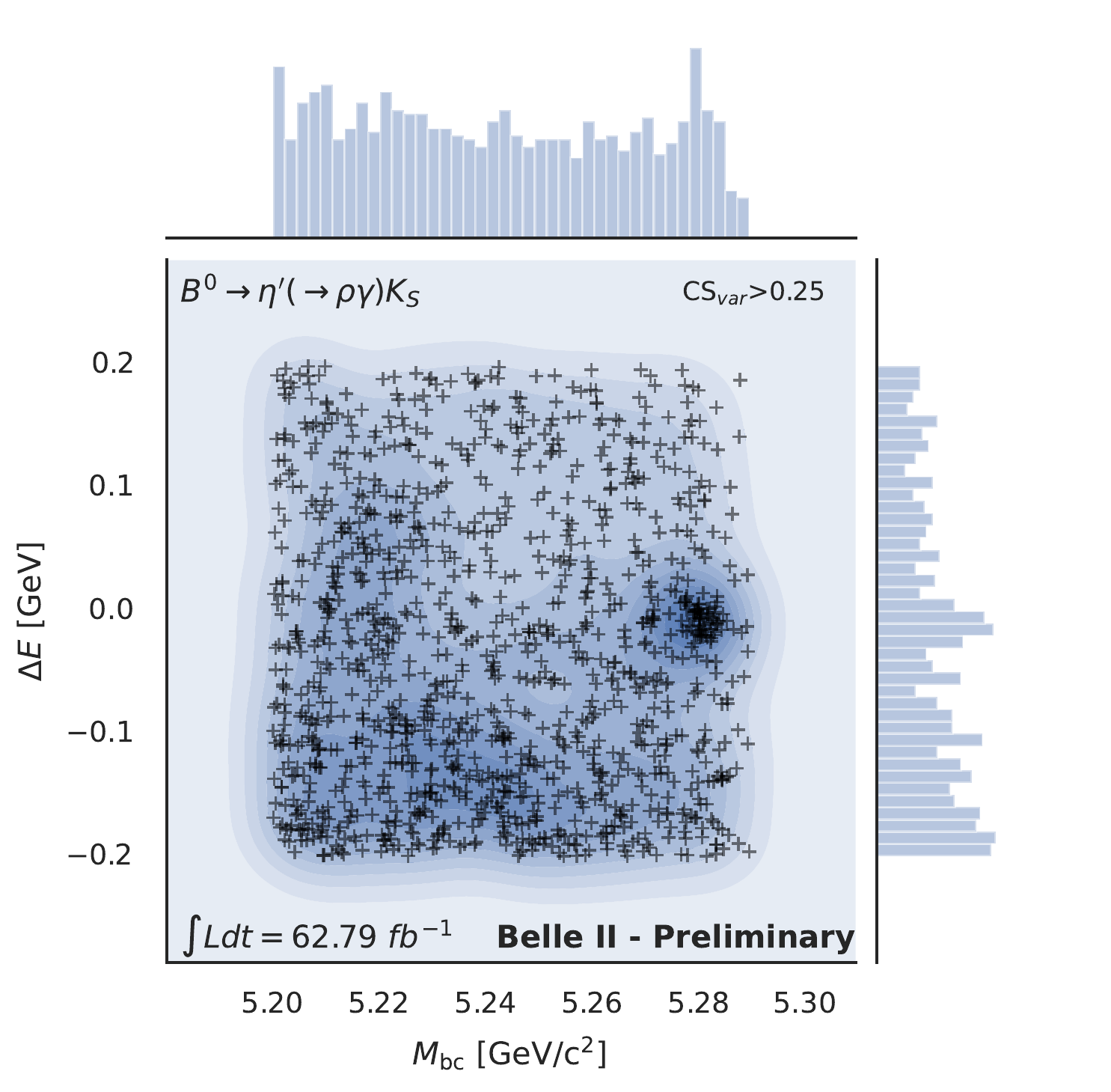}

    \includegraphics[width=.48\textwidth]{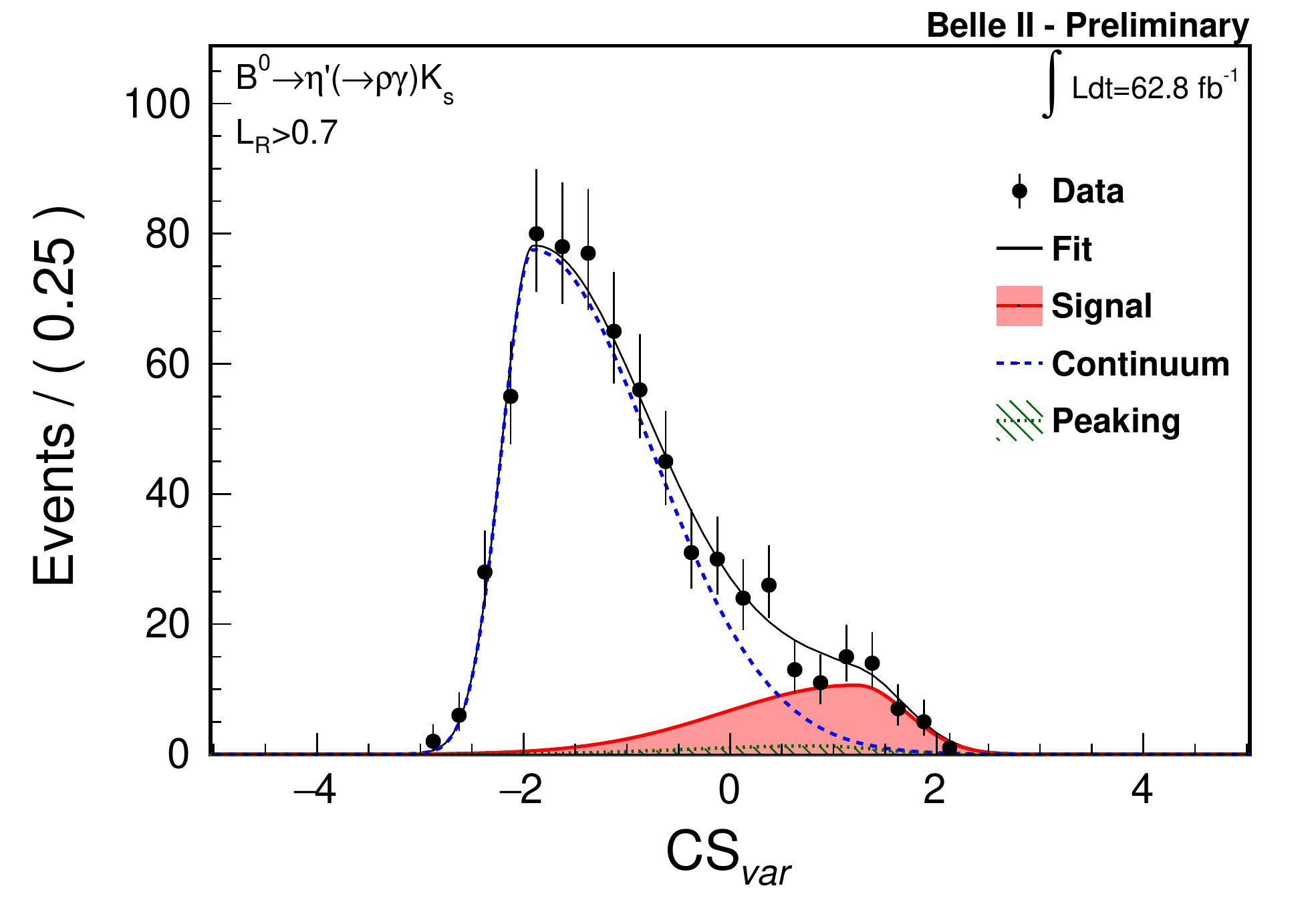}
    \includegraphics[width=.48\textwidth]{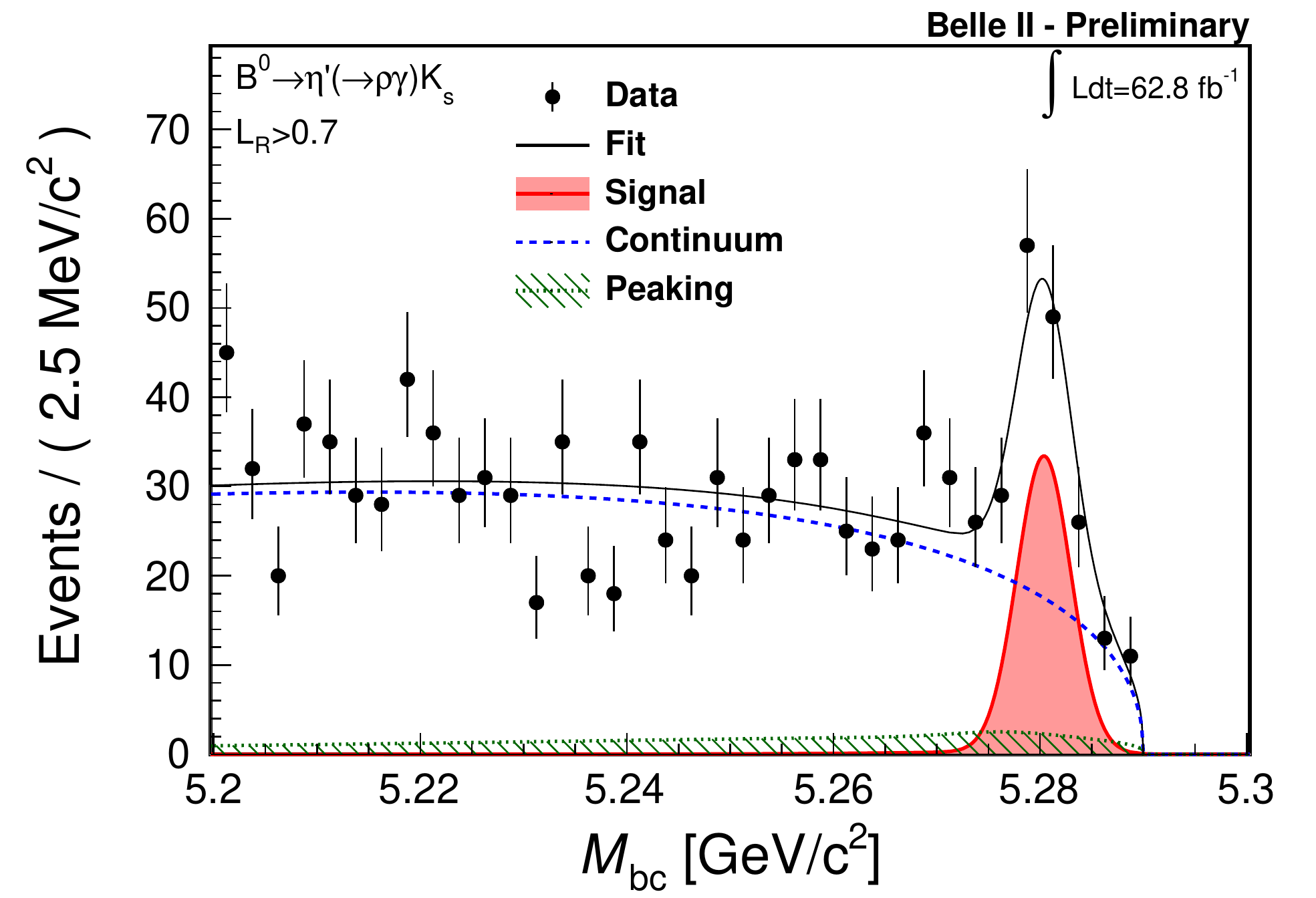}

    \caption{Distributions of $\Mbc$ and $\Delta{E}$, and continuum suppression discriminator for
        the signal-enriched region ($\mathcal{L}_R>0.7$), as well as $\Mbc$ versus $\Delta{E}$
        with the FoM-optimized \CS selection reported on the plot, for the channel 
        $\PBz\to\Petaprime\PKs$ with $\Petaprime\to\Prho\Pphoton$.
        Superimposed on the 1D distributions are
        the results of the extended ML fit as described in the text.
    }
       
    \label{fig:res_B0ch3}
\end{figure}

\begin{figure}[!htb]
    \includegraphics[width=.45\textwidth]{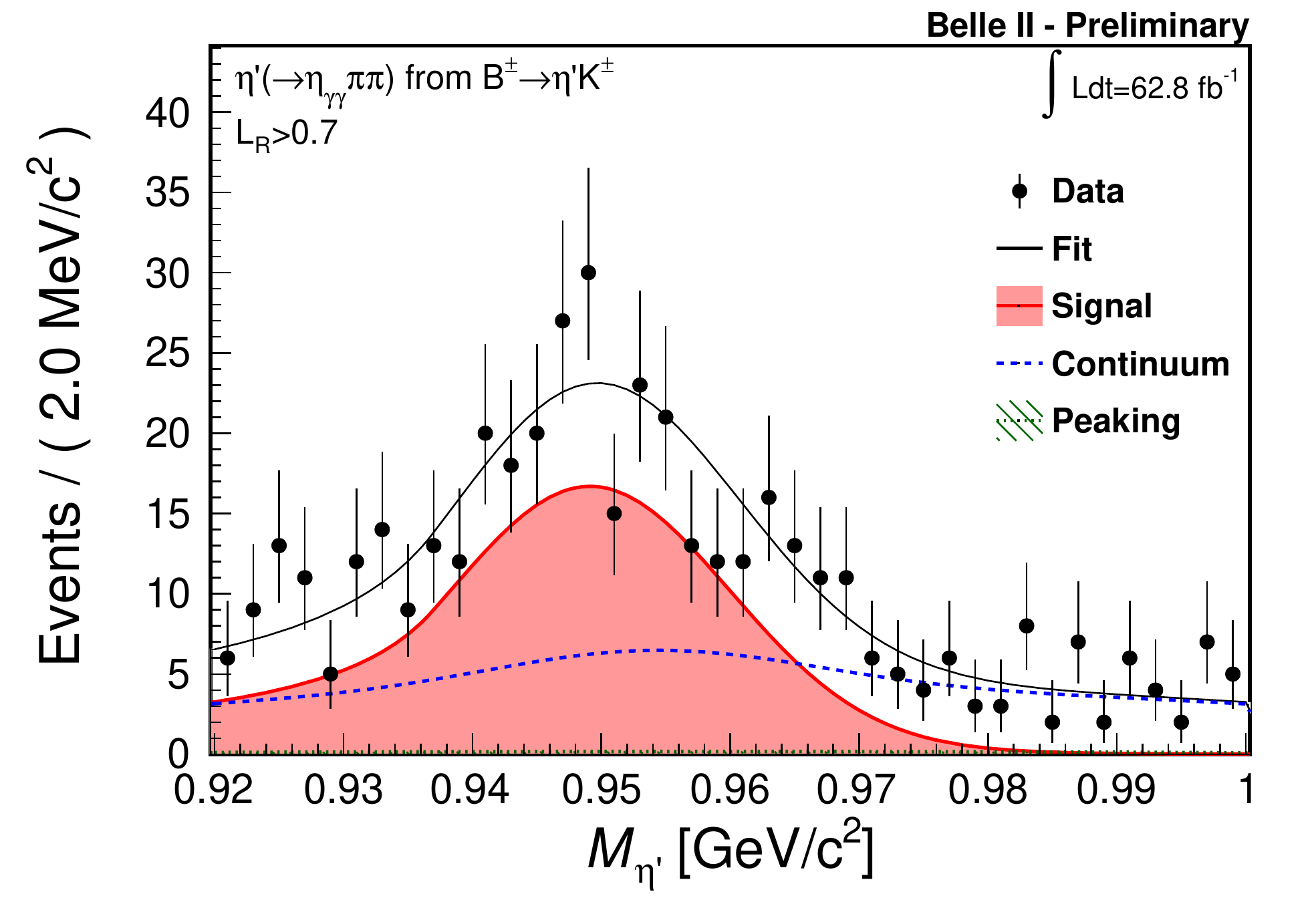}
    \includegraphics[width=.45\textwidth]{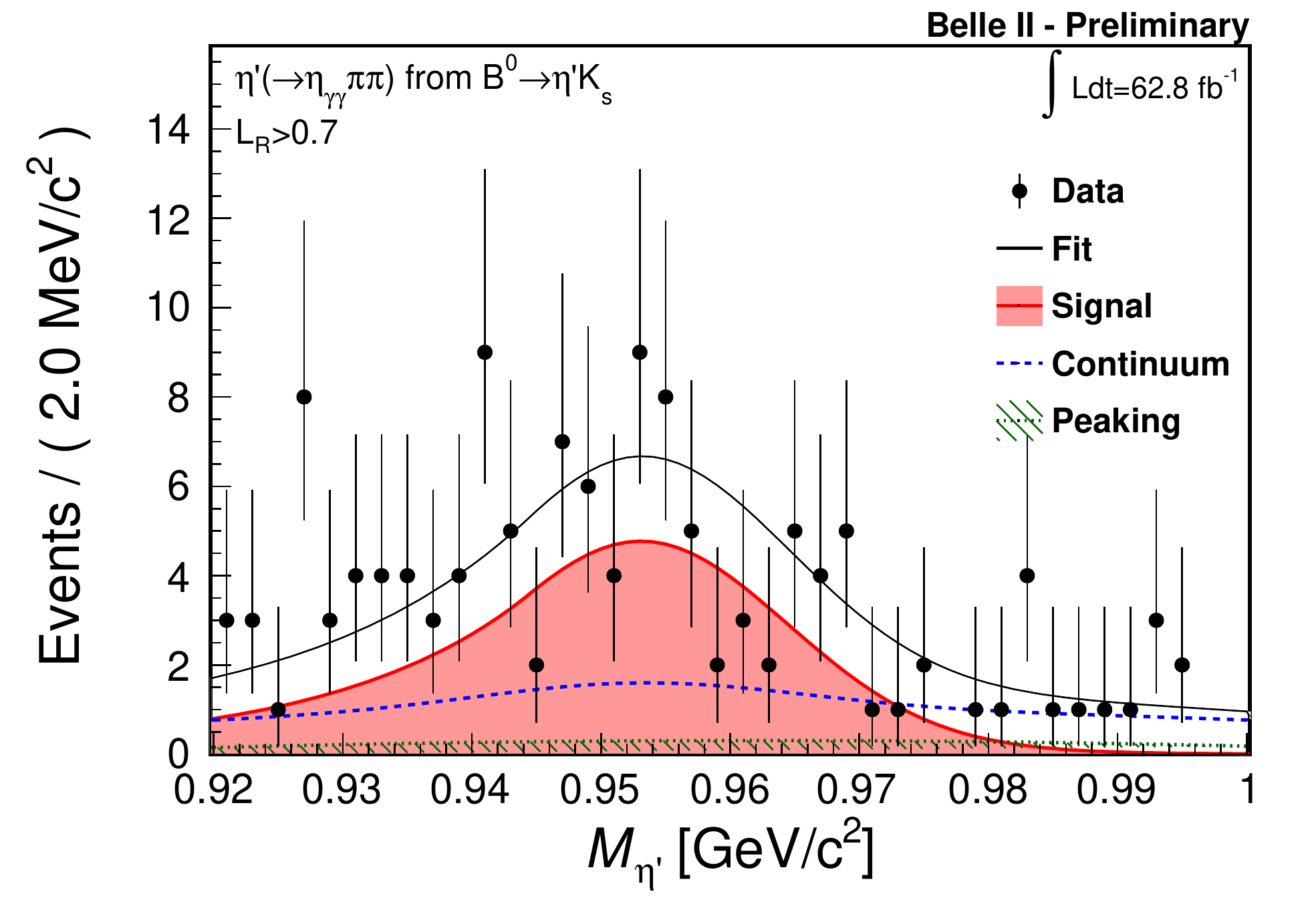}

    \includegraphics[width=.45\textwidth]{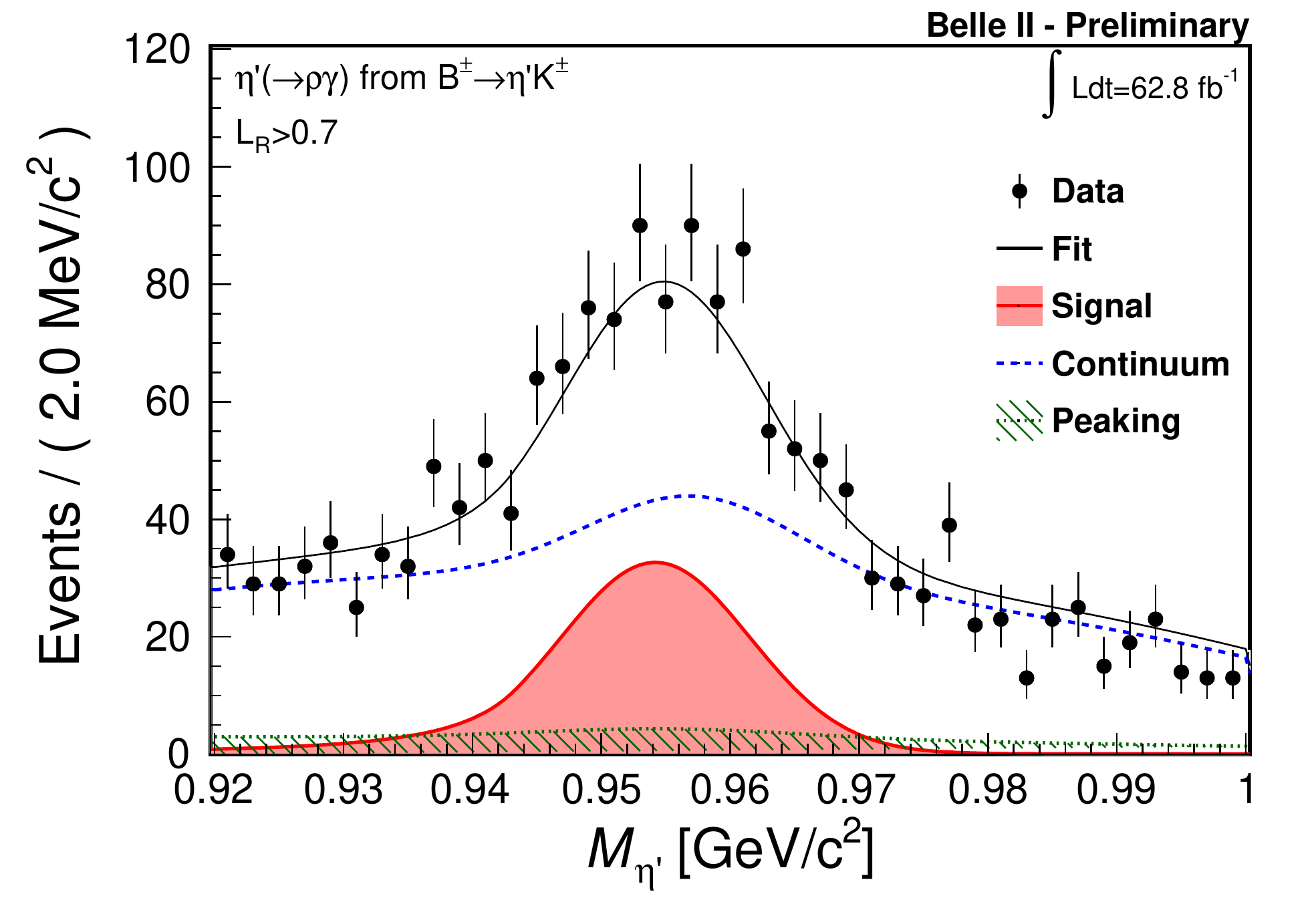}
    \includegraphics[width=.45\textwidth]{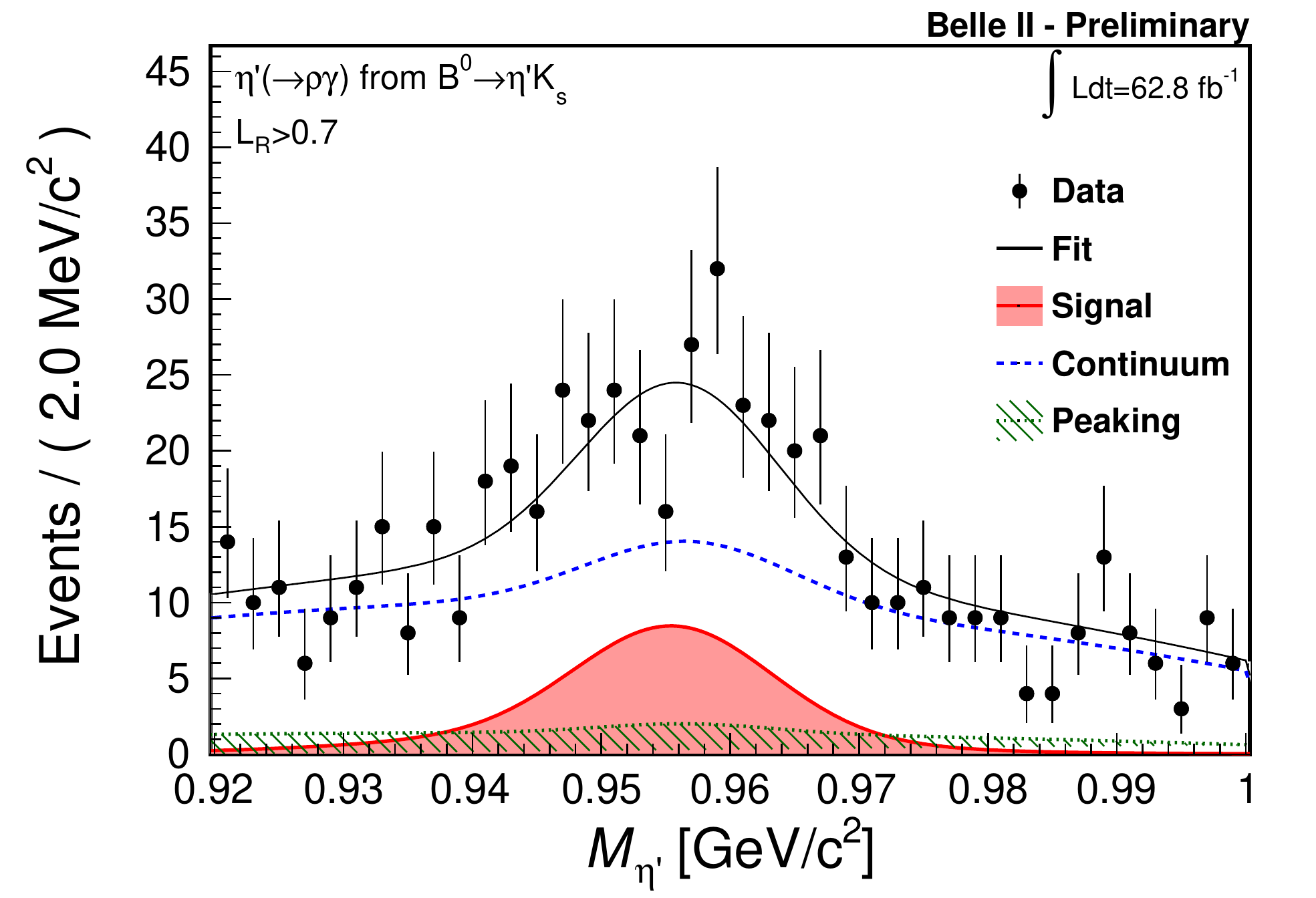}

    \caption{Distributions of invariant mass of \Petaprime, without any mass
        constraint, for the four channels in the signal-enriched region
        ($\mathcal{L}_R>0.7$).}
       
    \label{fig:Metap}
\end{figure}

The results of the ML fit on the 2019-2020 Belle II dataset at the \PUpsilonFourS are
reported in Table~\ref{tab:res} for the four channels analyzed.
In Fig.~\ref{fig:res_Bpch1}-\ref{fig:res_B0ch3}, the projections of the fit
results for the three observables are shown for data in a signal enriched
region, with a selection on signal versus background likelihood ratio 
($\mathcal{L}_R>0.7$).
The two dimensional distributions of \De vs \Mbc are shown with a selection on the continuum
suppression variable \CS, reported in the plots, which optimizes the figure of merit
$\text{FoM}=S/\sqrt{S+B}$, where $S$ and $B$ are the number of expected signal and
background events in the signal region, respectively.
Figure~\ref{fig:Metap} shows the distribution the  \Petaprime invariant mass
in the signal-enriched region for the four channels.
This variable has not been used for signal yield extraction.


The branching ratio is computed as:

\begin{equation}
    \mathcal{B}(B \to X) = \frac{N_{sig}}{2\cdot N(\BB) \cdot f_{00/+-} \cdot \varepsilon\B} \, ,
\end{equation}

\noindent{where} $N_{sig}$ is the yield for signal as returned by the ML fit, $N(\BB)$ is the
number of \BB pairs in the dataset, $f_{00/+-}$ is the fraction of
$\PBzero\APBzero$ and $\PBp\PBm$~\cite{Zyla:2020zbs}, respectively, and $\varepsilon\B$ is the
product of the signal reconstruction and selection efficiency and the branching
ratios of the daughter particles for the decay channels considered.

\begin{table}[!hbt]
    \caption{Summary of results for the four decay channels, corresponding to an integrated
        luminosity of $\L=62.8 \invfb$. Measured signal yields ($N_{sig}$),
        statistical significances ($sig.$), efficiencies ($\varepsilon$), total
        efficiencies including the secondary branching ratios ($\varepsilon\B$), and the
        measured \B are reported.
        The uncertainties are statistical, the second uncertainty in the
        last column (\B) is the systematic uncertainty.
    \label{tab:res}
    }
    \begin{tabular}{l@{\hskip 3mm}c@{\hskip 4mm}c@{\hskip 4mm}c@{\hskip 4mm}c@{\hskip 4mm}c@{\hskip 4mm}}
        \hline
        \hline
        Mode & $N_{sig}$ & $sig.$ & $\varepsilon(\%)$ & $\varepsilon\B(\%)$ & \B ($10^{-6}$) \\
        \hline
        $\PBpm\to\Petaprime(\to\Peta(\to\gamma\gamma)\Pgpp\Pgpm)\PKpm$  & $263~^{+18}_{-19}$        & $25.7$    & $31.7 \pm 0.03$  & $5.45$ & $63.9~^{+4.6}_{-4.4}\pm4.0$\\
        $\PBpm\to\Petaprime(\Prho(\to\Pgpp\Pgpm)\Pphoton)\PKpm$         & $335~^{+26}_{-25}$        & $22.2$    & $24.2 \pm 0.04$  & $7.05$ & $62.9~^{+4.8}_{-4.8}\pm5.5$\\

        \hline
        $\PBz\to\Petaprime(\to\Peta(\to\gamma\gamma)\Pgpp\Pgpm)\PKs$    & $80.0~^{+11.2}_{-10.4} $  & $13.8$    & $31.0 \pm 0.03$  & $1.80$ & $61.6~^{+8.6}_{-8.0}\pm3.9$\\
        $\PBz\to\Petaprime(\Prho(\to\Pgpp\Pgpm)\Pphoton)\PKs$           & $99.7~^{+14.2}_{-12.7} $  & $14.2$    & $23.6 \pm 0.04$  & $2.35$ & $58.5~^{+7.9}_{-7.4}\pm4.4$\\
        \hline
        \hline

    \end{tabular}
\end{table}

\section{Systematics uncertainties}
\label{sec:systematics}

The systematic uncertainties considered for this analysis are the following:

\begin{itemize}
    \item {\bf tracking efficiency}: we add 0.69\% for each charged track in the signal final state~\cite{Collaboration:2035};
    \item {\bf photon efficiency}: from a sample of
    $\Pelectron\Ppositron\to\Pmuon\APmuon\Pphoton$ events, the systematic uncertainties have been
    evaluated as a function of photon energy and polar angle $\theta$;
    \item {\bf \PKs reconstruction efficiency}: comparing data and simulation,
    we observed that the  ratio of \PKs reconstruction efficiency changes linearly as a
    function of the flight distance, so we applied an uncertainty of 0.31\% per
    cm of the average flight length, plus a 15\% uncertainty for the mis-modeling of
    material between second and third layer of SVD (10\% of candidates);
    \item {\bf PID}: a sample of $\PDstar\to\PD(\to\PK\Pgp)\Pgp_{soft}$ is used to compute
    the difference in efficiency for data and simulation as a function of
    particle momentum and $\cos\theta$ for pions and kaons\cite{Collaboration:2052};
    \item {\bf Continuum suppression modelling}: we repeat the ML fit using
    off-resonance data to obtain the pdf parameters for the continuum suppression
    discriminator, and use the difference in yield using the pdf from simulation as the uncertainty;
    \item {\bf SxF fraction}: we vary the fraction of SxF to truth matched
    signal found in simulation by $\pm50\%$ and assign the difference in yield
    as systematic uncertainty;
    \item {\bf N(\BB) yield}: an overall uncertainty of 1.4\% is taken as a systematic uncertainty, and
    includes the uncertainty on the cross section, integrated luminosity, and
    possible shift of collision energy~\cite{Collaboration:1997};
\end{itemize}

The systematic uncertainties for the four channels are summarized in Table~\ref{tab:syst}.

\begin{table}[hbt]
    \caption{Summary of systematics uncertainties (in \%) by category and channel.
        \label{tab:syst}}
        \begin{tabular}{l@{\hskip 3mm}c@{\hskip 4mm}c@{\hskip 4mm}c@{\hskip 4mm}c@{\hskip 4mm}}
        \hline
        \hline
            \hfill{Channel} &  {$\PBpm\to\Petaprime\PKpm$} & {$\PBz\to\Petaprime\PKs$} &  {$\PBpm\to\Petaprime\PKpm$} & {$\PBz\to\Petaprime\PKs$} \\
            {Source}&  \multicolumn{2}{c}{$\Petaprime\to\Peta\Pgpp\Pgpm$} & \multicolumn{2}{c}{$\Petaprime\to\Prho\Pphoton$} \\
            \hline
            Tracking efficiency   & 2.1 & 2.8 & 2.1 & 2.8 \\
            Photon efficiency     & 0.5 & 0.5 & 0.5 & 0.5 \\
            \PKs efficiency       & -   & 4.5 & -   & 4.5 \\
            \Pgppm PID            & -   & -   & 2.4 & 2.4 \\
            \PKpm PID             & 2.5 & -   & 2.5 & -   \\
            Cont. supp. modelling & 5.0 & 1.0 & 5.5 & 2.3 \\
            SxF fraction          & 2.6 & 1.8 & 5.9 & 3.2 \\ 
            N(\BB) & \multicolumn{4}{c}{1.4} \\
            \hline
            Total                 & 6.6 &  5.9 & 9.1 & 7.2\\
        \hline
        \hline
  \end{tabular}
\end{table}

\section{Conclusions}
\label{sec:conclusions}

In this paper, we described the rediscovery of the $\PB\to\Petaprime\PK$ hadronic-penguin mediated
decays, both for charged and neutral \PB mesons.
The full Belle II luminosity collected up to summer 2020 has been used,
corresponding to $L=62.8\invfb$ and $68.2\times10^{6}$ \BB pairs.
The signal yield is extracted via a maximum likelihood fit to the data, and has been validated
on simulation, as well as on off-resonance data.
The results, reported in Table~\ref{tab:resB}, are in good agreement with world averages~\cite{Zyla:2020zbs}.
The signal yield per $10^{6}~\BB$ is similar to that reported by
BaBar~\cite{Aubert:2009yx}, and almost a factor two larger that that of
Belle~\cite{Schumann:2006bg}, partially thanks to the absence of selection on
continuum suppression variable.
The next step will be to use the future large data sample collected at Belle II for a full time
dependent CP violation analysis.

\begin{table}[!hbt]
    \caption{Summary of results on branching ratios obtained in this analysis, and comparison with world averages. 
        \label{tab:resB}}
    \begin{tabular}{l@{\hskip 3mm}c@{\hskip 4mm}c@{\hskip 4mm}}
        \hline
        \hline
        & {This analysis} & {World average~\cite{Zyla:2020zbs}} \\
        Channel & \multicolumn{2}{c}{\B($\times10^{6})$} \\
        \hline
        $\PBpm\to\Petaprime\PK$  & $63.4~^{+3.4}_{-3.3}\stat\pm3.4\syst$ & $70.4\pm2.5$  \\
        $\PBz\to\Petaprime\PKz$  & $59.9~^{+5.8}_{-5.5}\stat\pm2.7\syst$ & $66\pm4$ \\
        \hline
        \hline

    \end{tabular}
\end{table}

\section*{Acknowledgements}
We thank the SuperKEKB group for the excellent operation of the
accelerator; the KEK cryogenics group for the efficient
operation of the solenoid; the KEK computer group for
on-site computing support; and the raw-data centers at
BNL, DESY, GridKa, IN2P3, and INFN for off-site computing support.
This work was supported by the following funding sources:
Science Committee of the Republic of Armenia Grant No. 20TTCG-1C010;
Australian Research Council and research grant Nos.
DP180102629, 
DP170102389, 
DP170102204, 
DP150103061, 
FT130100303, 
FT130100018,
and
FT120100745;
Austrian Federal Ministry of Education, Science and Research,
Austrian Science Fund No. P 31361-N36, and
Horizon 2020 ERC Starting Grant no. 947006 ``InterLeptons''; 
Natural Sciences and Engineering Research Council of Canada, Compute Canada and CANARIE;
Chinese Academy of Sciences and research grant No. QYZDJ-SSW-SLH011,
National Natural Science Foundation of China and research grant Nos.
11521505,
11575017,
11675166,
11761141009,
11705209,
and
11975076,
LiaoNing Revitalization Talents Program under contract No. XLYC1807135,
Shanghai Municipal Science and Technology Committee under contract No. 19ZR1403000,
Shanghai Pujiang Program under Grant No. 18PJ1401000,
and the CAS Center for Excellence in Particle Physics (CCEPP);
the Ministry of Education, Youth and Sports of the Czech Republic under Contract No.~LTT17020 and 
Charles University grants SVV 260448 and GAUK 404316;
European Research Council, 7th Framework PIEF-GA-2013-622527, 
Horizon 2020 ERC-Advanced Grants No. 267104 and 884719,
Horizon 2020 ERC-Consolidator Grant No. 819127,
Horizon 2020 Marie Sklodowska-Curie grant agreement No. 700525 `NIOBE,' 
and
Horizon 2020 Marie Sklodowska-Curie RISE project JENNIFER2 grant agreement No. 822070 (European grants);
L'Institut National de Physique Nucl\'{e}aire et de Physique des Particules (IN2P3) du CNRS (France);
BMBF, DFG, HGF, MPG, and AvH Foundation (Germany);
Department of Atomic Energy under Project Identification No. RTI 4002 and Department of Science and Technology (India);
Israel Science Foundation grant No. 2476/17,
United States-Israel Binational Science Foundation grant No. 2016113, and
Israel Ministry of Science grant No. 3-16543;
Istituto Nazionale di Fisica Nucleare and the research grants BELLE2;
Japan Society for the Promotion of Science,  Grant-in-Aid for Scientific Research grant Nos.
16H03968, 
16H03993, 
16H06492,
16K05323, 
17H01133, 
17H05405, 
18K03621, 
18H03710, 
18H05226,
19H00682, 
26220706,
and
26400255,
the National Institute of Informatics, and Science Information NETwork 5 (SINET5), 
and
the Ministry of Education, Culture, Sports, Science, and Technology (MEXT) of Japan;  
National Research Foundation (NRF) of Korea Grant Nos.
2016R1\-D1A1B\-01010135,
2016R1\-D1A1B\-02012900,
2018R1\-A2B\-3003643,
2018R1\-A6A1A\-06024970,
2018R1\-D1A1B\-07047294,
2019K1\-A3A7A\-09033840,
and
2019R1\-I1A3A\-01058933,
Radiation Science Research Institute,
Foreign Large-size Research Facility Application Supporting project,
the Global Science Experimental Data Hub Center of the Korea Institute of Science and Technology Information
and
KREONET/GLORIAD;
Universiti Malaya RU grant, Akademi Sains Malaysia and Ministry of Education Malaysia;
Frontiers of Science Program contracts
FOINS-296,
CB-221329,
CB-236394,
CB-254409,
and
CB-180023, and SEP-CINVESTAV research grant 237 (Mexico);
the Polish Ministry of Science and Higher Education and the National Science Center;
the Ministry of Science and Higher Education of the Russian Federation,
Agreement 14.W03.31.0026;
University of Tabuk research grants
S-0256-1438 and S-0280-1439 (Saudi Arabia);
Slovenian Research Agency and research grant Nos.
J1-9124
and
P1-0135; 
Agencia Estatal de Investigacion, Spain grant Nos.
FPA2014-55613-P
and
FPA2017-84445-P,
and
CIDEGENT/2018/020 of Generalitat Valenciana;
Ministry of Science and Technology and research grant Nos.
MOST106-2112-M-002-005-MY3
and
MOST107-2119-M-002-035-MY3, 
and the Ministry of Education (Taiwan);
Thailand Center of Excellence in Physics;
TUBITAK ULAKBIM (Turkey);
Ministry of Education and Science of Ukraine;
the US National Science Foundation and research grant Nos.
PHY-1807007 
and
PHY-1913789, 
and the US Department of Energy and research grant Nos.
DE-AC06-76RLO1830, 
DE-SC0007983, 
DE-SC0009824, 
DE-SC0009973, 
DE-SC0010073, 
DE-SC0010118, 
DE-SC0010504, 
DE-SC0011784, 
DE-SC0012704, 
DE-SC0021274; 
and
the Vietnam Academy of Science and Technology (VAST) under grant DL0000.05/21-23.

\bibliography{references}

\end{document}